\documentclass[twocolumn]{autart}    

\usepackage{graphicx}          

\usepackage{cite}
\usepackage{amsmath,amssymb,amsfonts}
\usepackage{graphicx}
\usepackage{textcomp}
\usepackage{arydshln}

\usepackage{natbib}

\usepackage[usenames,dvipsnames]{color}

\definecolor{V1R}{RGB}{200,100,0}

\usepackage{graphicx}
\usepackage[dvips]{epsfig} 
\usepackage{epstopdf}
\usepackage{amsmath, bm}
\usepackage{caption}
\usepackage{enumerate}
\usepackage{booktabs}
\usepackage{setspace}
\usepackage{stfloats}
\usepackage{amsfonts}
\usepackage{threeparttable}

\usepackage[colorlinks,
linkcolor=cyan,
anchorcolor=cyan,
citecolor=cyan,
hyperfigures= TRUE,
]{hyperref}

\newtheorem{define}{Definition}
\newtheorem{lemma}{Lemma}
\newtheorem{proposition}{Proposition}
\newtheorem{remark}{Remark}
\newtheorem{theorem}{Theorem}
\newtheorem{corl}{Corollary}

\usepackage{url}

\makeatletter

\newcommand{\Rmnum}[1]{\expandafter\@slowromancap\romannumeral #1@}
\makeatother

\def\QEDopen{{\setlength{\fboxsep}{0pt}\setlength{\fboxrule}{0.2pt}\fbox{\rule[0pt]{0pt}{1.3ex}\rule[0pt]{1.3ex}{0pt}}}} 

\usepackage{color}
\usepackage{xcolor}

\usepackage{subfig}
\usepackage{algorithm,algorithmic}
\usepackage{float}
\usepackage{easyReview}

\begin{document}

\begin{frontmatter}

\title{Information-triggered Learning with Application to Learning-based Predictive Control} 


\author[MIIT]{Kaikai Zheng}\ead{kaikai.zheng@bit.edu.cn},    
\author[MIIT]{Dawei Shi}\ead{daweishi@bit.edu.cn},               
\author[Hirche]{Sandra Hirche}\ead{hirche@tum.de},  
\author[Shi]{Yang Shi}\ead{yshi@uvic.ca}  

\address[MIIT]{School of Automation, Beijing Institute of Technology, Beijing 100081, China}  
\address[Hirche]{Chair of Information-oriented Control, Technical
	University of Munich, Barer Strasse 21, 80333, Munich, Germany}             
\address[Shi]{Department of Mechanical Engineering, Faculty of Engineering, University of Victoria, Victoria, BC V8N 3P6, Canada}        


\begin{abstract}                          
	Learning-based control has attracted significant attention in recent years, especially for plants that are difficult to model based on first-principles. A key issue in learning-based control is how to make efficient use of data as the abundance of data becomes overwhelming. To address this issue, this work proposes an information-triggered learning framework and a corresponding learning-based controller design approach with guaranteed stability. Specifically, we consider a linear time-invariant system with unknown dynamics. A set-membership approach is introduced to learn a parametric uncertainty set for the unknown dynamics. Then, a data selection mechanism is proposed by evaluating the incremental information in a data sample, where the incremental information is quantified by its effects on shrinking the parametric uncertainty set. Next, after introducing a stability criterion using the set-membership estimate of the system dynamics, a robust learning-based predictive controller (LPC) is designed by minimizing a worst-case cost function. The closed-loop stability of the LPC equipped with the information-triggered learning protocol is discussed within a high-probability framework. Finally, comparative numerical experiments are performed to verify the validity of the proposed approach.
\end{abstract}

\end{frontmatter}

\section{Introduction}
Recent developments in advanced sensing and communication technology offer an increasing volume of data for control systems design, which accelerates the advancements on the methodological front of learning-based control \citep{de2019formulas, van2020data, woo2021overview, markovsky2023data}. Among others, data efficiency is an important issue that needs to be addressed as the abundance of data becomes overwhelming in control systems. In this work, we aim to introduce a solution by online selecting the data that contains information of the system dynamics within a set-membership framework, and integrating the data selection scheme with learning-based predictive controller (LPC) design.

Set-membership learning methods offer a systematic framework to learn system dynamics and evaluate the learning performance \citep{milanese2004set, milanese2013bounding}. 
For linear systems with Gaussian noise, the authors in \citet{umenberger2019robust} proposed a set-membership learning method to estimate a credibility region of unknown parameters. 
Similarly, an uncertainty set was learned in \cite{dean2020sample} to bound the error in parameter estimation procedures used in the design of a robust controller.
An admissible set of system parameters was formulated as quadratic matrix inequality (QMI) for linear systems with bounded noise in \cite{van2020data}. 
Set-membership learning methods are also proved useful in learning nonlinear systems \citep{karimshoushtari2020design}.
For other methods of set-membership learning, the interested readers can refer to \cite{ozay2015set,lauricella2020set} and references therein.

The focus of this work is also closely related to LPC, which features the adoption of a learning-based prediction model in the constrained optimization problem solved to obtain the controller output at each sampling instant. Several LPC methods have been proposed in the literature by exploiting the advantages of different data-driven models \citep{korda2018linear,verheijen2023handbook}.  
Based on the renowned fundamental lemma \cite{willems2005note}, a Data-Enabled Predictive Control (DeePC) was first proposed for linear systems \citep{coulson2019data} and then generalized to various types of settings, including noise-corrupted data \citep{coulson2019regularized,coulson2021distributionally}, nonlinear systems \citep{elokda2021data,huang2023robust}, online learning and control \citep{schmitt2023data}.
Gaussian processes were also employed in the design of LPC \citep{bradford2020stochastic}.
For instance, the LPC in \cite{hewing2019cautious} learns the additive nonlinear model mismatch using a Gaussian process.
In addition, set-membership models were also discussed for LPC \citep{tanaskovic2014adaptive,lorenzen2019robust}, and the min-max framework is a useful tool for designing corresponding robust predictive controllers \citep{xie2024data}. In min-max predictive control, the worst-case cost function is minimized over a set of admissible disturbances or parameters. Several approaches were proposed for solving min-max predictive control problems, including reformulation \citep{diehl2007formulation}, robust dynamic programming \citep{bjornberg2006approximate}, scenario approach \citep{1632303}, and constraints relaxation \citep{hu2022handling}.
However, with the accumulation of incoming data during the operation of the control system, an important question to answer is when to update the prediction model for improved data efficiency (and consequently reduced computation complexity) while guaranteeing control performance.

Event-triggered learning (ETL) provides an effective remedy to overcome this issue by learning only when certain pre-specified conditions are violated.
ETL can improve computational efficiency by reducing the complexity of non-parametric models or by decreasing the updating frequency of parametric models.
For non-parametric models, more available data usually leads to the enlarged computational burden, which can be alleviated using ETL by learning from selected data samples \citep{he2022learning,zheng2022non}.
Specifically, for Gaussian processes, data samples were selected according to model uncertainty criterion \citep{umlauft2019feedback,lederer2021gaussian,jiao2022backstepping} or associated entropy-based criterion \citep{umlauft2020smart}.
For parametric models, although the complexity of the model does not change with the amount of training data, the frequent model updates would lead to increased computational burden. 
Several ETL approaches were therefore proposed for parametric models by identifying the change in system dynamics and adjusting the model only when necessary \citep{Trimpe2020ETL}. 
For linear systems with intermittent communication processes, the authors in \cite{Trimpe2020ETLTAC} proposed two learning mechanisms triggered by the distribution and expectation of inter-communication time, respectively. 
Despite the advances in ETL, it is not clear regarding how to quantify and evaluate the importance of the incoming data samples online in exploiting the structural properties of the underlying system dynamics, which motivates the investigation in this work.

The objective of our work is to introduce a systematic data selection method and an LPC design approach with guaranteed stability. 
Specifically, we consider a linear time-invariant (LTI) system with unknown dynamics and disturbances, and the dataset is updated in an \textit{information-triggered learning (ITL)} fashion. 
Here the triggering condition is designed using the incremental information contained in an incoming data sample, which is quantified by its effect on shrinking the set of potential dynamics that compatible with available data.
We consider the scenario that only limited prior knowledge of the disturbance is available, which is characterized with a form of generic concentration inequality.
A few challenges, however, need to be addressed to enable the design of the ITL protocol and the corresponding LPC. 
First, the effect of the unknown disturbance with the adopted prior knowledge blurs the characteristics of the system dynamics, which makes it more challenging to ensure the learning performance.
Second, since the information of a data sample depends on the unknown system dynamics and the data recorded previously, it is challenging to design an online data selection protocol that identifies a slim dataset to support a learning algorithm with ensured stability.
In addition, the inferred system dynamics in this work are represented using a set-membership method and are updated intermittently, which adds to the difficulties in LPC design and the corresponding stability analysis.
The main contributions of this work are summarized as follows:


\begin{enumerate}
	\item Compared with the existing literature that considered bounded deterministic disturbance or Gaussian disturbance, a set-membership learning method for linear systems is proposed in the form of a QMI for disturbances modeled by a generic concentration inequality.
y.
	Several matrix parameters are designed for the set-membership learning method, which ensures that the learned parametric uncertainty set contains the unknown parameters with a high-probability. 
	\item Utilizing the proposed set-membership learning method, an ITL mechanism is introduced by quantifying the information of a new data sample.
	Specifically, the incremental information contained in an incoming data sample is quantified in terms of its effect on shrinking the set of system dynamics compatible with available data. It is proved that, with a predefined probability, the Lebesgue measure of the learned parametric uncertainty set decreases exponentially under the ITL mechanism.
	The special cases of bounded deterministic disturbances and stochastic disturbances with covariance information are discussed.
	\item The applicability of the ITL approach to LPC design is demonstrated. To do this, a constrained optimization problem is constructed within a min-max framework using the proposed ITL schemes. The proposed control approach features the design of a robust LPC that intermittently updates its prediction model only when the incoming data can help reduce the uncertainty of the learned system dynamics. The closed-loop stability is guaranteed in the sense of high-probability by designing a data-based linear matrix inequality (LMI).
\end{enumerate}

The remainder of this paper is organized as follows. 
Section \ref{sec:pf} presents the main problem considered in our work, useful definitions and lemmas are also introduced. 
The main theoretic results on set-membership learning, ITL, and their application to LPC design are presented in Section \ref{sec:Result}.
Moreover, implementation issues, numerical verification, and comparison results are comprehensively presented in Section \ref{sec:nbex}, followed by the concluding remarks in Section \ref{sec:conc}.

\textbf{Notation.~}In this work, ${\text{diag}\{x_1,\ldots,x_n\}}$ represents a diagonal matrix with diagonal elements $\{x_1,\ldots,x_n\}$, and $I_n$ represents an $n$-dimensional identity matrix. For matrices $A$ and $B$, $A\succ B$ and $A\succeq B$ denote that the matrix $A-B$ is positive definite and positive semi-definite, respectively. Moreover, $A\npreceq B$ denotes that the matrix $B-A$ is not positive semi-definite. For a random variable, $\mathbb{P}[\cdot]$ and $\mathbb{E}[\cdot]$ represent the probability and the expectation, respectively. The eigenvalues of a matrix $A\in\mathbb{R}^{n\times n}$ are written as ${\bm\lambda}(A)=[\lambda_1(A),\lambda_2(A),\ldots,\lambda_n(A)]$ with $|\lambda_1(A)|\geq|\lambda_2(A)|\geq\ldots\geq|\lambda_n(A)|$, and the Moore-Penrose inverse of $A$ is denoted as $A^\dagger$. The cardinality of a set $\mathcal{D}$ is denoted as $\|\mathcal{D}\|$. Moreover, $|A|$ and ${\rm Tr}(A)$ denote the determinant and the trace of matrix $A$, respectively. Additionally, for simplicity, $A\star^{\rm T}$ is used to denote $AA^{\rm T}$.

\section{Problem Formulation}\label{sec:pf}
Consider a linear time-invariant system
\begin{align}\label{eq:sys}
	{\bm x}(k+1)&=A_*{\bm x}(k)+B_*{\bm u}(k)+{\bm w}(k),
\end{align}%
where ${\bm x}(k)\in\mathbb{R}^{n_x\times 1}$ is the state, ${\bm u}(k)\in\mathbb{R}^{n_u\times 1}$ is the input, ${\bm w}(k)$ is the disturbance, and $A_*\in\mathbb{R}^{n_x\times n_x}, B_*\in\mathbb{R}^{n_x\times n_u}$ are unknown system parameters.

The data samples collected up to time step $k$ are represented as $\mathcal{D}(k)$, and the index set of the dataset is denoted as $\mathcal{R}(k)$. Therefore, the dataset $\mathcal{D}(k)$ can be expressed as
\begin{align*}
	\mathcal{D}(k):=\{d(i)|i\in\mathcal{R}(k)\},~~
	\mathcal{R}(k):=\{r_1,r_2,\ldots,r_{n(k)}\},
\end{align*}
where $d(i)=\{\bm{x}(i),\bm{u}(i),\bm{x}(i+1)\}$, and $n(k)=\|\mathcal{R}(k)\|$ is the number of the data samples in $\mathcal{D}(k)$. Moreover, the data recorded in the dataset $\mathcal{D}(k)$ can be written in a compact form as
\begin{align}
	X_{_-}(k)&\!=\![\bm{x}(r_1), \bm{x}(r_2), \ldots, \bm{x}(r_{n(k)})],\notag\\
	U_{_-}(k)&\!=\![\bm{u}(r_1), \bm{u}(r_2), \ldots, \bm{u}(r_{n(k)})],\notag\\
	W_{_-}(k)&\!=\![\bm{w}(r_1), \bm{w}(r_2), \ldots, \bm{w}(r_{n(k)})],\notag\\
	X_{_+}(k)&\!=\![\bm{x}(r_1\!+\!1), \bm{x}(r_2\!+\!1), \ldots, \bm{x}(r_{n(k)}\!+\!1)].\notag
\end{align}
\vspace{-1cm}

In this work, we mainly consider the scenario where the disturbance ${\bm w}(k), k\in\mathbb{N}$ is an independent, identically distributed (i.i.d.) stochastic process with zero mean. Note that the assumption of zero mean is commonly adopted in existing literature \citep{piga2017direct,abraham2019active,baggio2021data}; by introducing the bias values of the system state, a nonzero mean disturbance can be further transformed into the equivalent scenario of a zero mean disturbance.
Moreover, we assume that the disturbances in matrix $W_-(k)$ satisfy the following property:
\begin{align}\label{eq:introdelta}
	\mathbb{P}[W_{-}(k)W_{-}^{\rm T}(k)\preceq\Phi_1(n(k),\delta)]\geq\delta,
\end{align}
where $\Phi_1(n(k),\delta)$ is a positive semi-definite matrix.

\begin{remark} Inequality \eqref{eq:introdelta} is normally called a concentration inequality, which holds for different kinds of disturbances. 
	For the i.i.d. disturbances in matrix $W_-(k)$, the specific parameter $\Phi_1(n(k),\delta)$ is related with the number of the considered data samples $n(k)$ and the probability $\delta$. In practice, the matrix parameter $\Phi_1(n(k),\delta)$ can be obtained according to various types of prior knowledge of the disturbance, which allows the application of the proposed results to different scenarios.
	After completing the design of the ITL mechanism and discussing its related properties, several specific examples for $\Phi_1(\cdot)$ will be briefly discussed in Section \ref{sec:conineq}.
\end{remark}

To learn unknown system dynamics, we define the admissible set of system parameters $(A,B)$ that are compatible to the dataset $\mathcal{D}(k)$ as
\begin{align}
	&\Gamma_\delta(\mathcal{D}(k))\label{eq:defGamma}\\
	:=&\Big\{(A,B)\Big{|}X_+(k)=AX_-(k)+BU_-(k)+{W}_-(k)~\text{holds}\notag\\
	&~~~~~~~~~~~~~	\text{for some}~{W}_-(k)~\text{satisfying}~\eqref{eq:introdelta}\Big\}.\notag
\end{align}
Let $\mathfrak{V}(\Gamma_\delta(\mathcal{D}(k)))$ be the Lebesgue measure of the set $\Gamma_\delta(\mathcal{D}(k))$ (see, e.g., \cite{boyd2004convex} for the calculation of the Lebesgue measure of a set). With this notation, we quantify the contribution of a new data sample $d(k)$ on reducing the Lebesgue measure of the set-membership estimate $\Gamma_\delta(\mathcal{D}(k-1))$, which is called quantitative incremental information in this work and is denoted as $\mathcal{I}_{\rm QII}(k)$.

\begin{define}\label{def:epinnovative}
	Consider a set $\Gamma_\delta(\mathcal{D}(k-1))$ defined in \eqref{eq:defGamma} and a new data sample $d(k)$. The quantitative incremental information of the data sample $d(k)$ is defined as
	\begin{align}
		\mathcal{I}_{\rm QII}(k)
		:=1-\frac{\mathfrak{V}\left(\Gamma_\delta\left({\mathcal{D}}(k-1)\cup d(k)\right)\right)}{\mathfrak{V}\big(\Gamma_\delta(\mathcal{D}(k-1))\big)}.\notag
	\end{align}
\end{define}
With the above descriptions, this work introduces an online data-selection and control-relevant information-triggered learning approach by evaluating the quantitative incremental information of a data sample. Specifically, the following questions will be investigated:
\begin{itemize}
	\item How to parameterize the set $\Gamma_\delta(\mathcal{D}(k))$ to facilitate the evaluation of $\mathfrak{V}(\Gamma_\delta(\mathcal{D}(k)))$?
	\item How to design an ITL mechanism to balance the learning performance and data efficiency?
	\item How to design an LPC that equips ITL to ensure the closed-loop stability?
\end{itemize}

\section{Main Results}\label{sec:Result}
This section presents the main theoretic developments of the proposed ITL mechanism and its application to LPC design.
We first introduce the proposed ITL approach and provide the corresponding convergence analysis. Then we discuss the parameterization of the concentration inequality (2) for two special cases of disturbances, including bounded deterministic disturbances and stochastic disturbances with known covariance. Finally, we show the applicability of the ITL scheme to a min-max predictive control approach.

\subsection{Set-membership Learning}\label{sec:etl}
In this section, we develop a probabilistic set-membership learning approach compatible with the dataset $\mathcal{D}(k)$ and the concentration inequality \eqref{eq:introdelta}. 		
To enable the design of the set-membership learning, we parameterize the set $\Gamma_\delta(\mathcal{D}(k))$ as follows:
\begin{align}\label{eq:defGama}
	&\Gamma_\delta(\mathcal{D}(k))\!=\!\Gamma(\Psi(n(k),\delta),\mathcal{D}(k))\\
	:=&\!\left\{(A,B)\left|Z(A,B){\Psi}(n(k),\delta)Z^{\rm T}(A,B)\succeq 0\right.\!\right\},\notag
\end{align}
where $Z(A,B):=[I_{n_x}~A~B]$ is a function and ${\Psi}(n(k),\delta)$ is a matrix parameter defined as 
\begin{align}
	&{\Psi}(n(k),\delta):=\Xi(k)\tilde{\Psi}(n(k),\delta)\Xi^{\rm T}(k),\label{eq:tPsi}\\
	&\Xi(k)\!:=\!\left[\!\!\begin{array}{cc}
		I_{n_x} & X_{_+}(k)\\{\bm 0}&-X_{_-}(k)\\{\bm 0} &-U_{_-}(k)
	\end{array}\!\!\right],
	\tilde{\Psi}(n(k),\delta)\!:=\!\left[\!\!\begin{array}{cc}
		{\Phi}_1(n(k),\delta)& {\bm 0}\\{\bm 0}&-I
	\end{array}\!\!\right].\label{eq:Phi2def}
\end{align}
In the next result, we show that the learned parametric uncertainty set $\Gamma({\Psi}(n(k),\delta),\mathcal{D}(k))$ contains unknown	parameters $(A_*,B_*)$ with a high-probability.

\begin{proposition}\label{thm:Pab}
	Consider system \eqref{eq:sys} and an available dataset $\mathcal{D}(k)$. If the set $\Gamma_\delta(\mathcal{D}(k))$ is parameterized as \eqref{eq:defGama}-\eqref{eq:Phi2def} and \eqref{eq:introdelta} holds, then the system matrices $A_*,B_*$ satisfy
	\begin{align}\label{eq:AtBt}
		\mathbb{P}\left[(A_*,B_*)\in \Gamma({\Psi}(n(k),\delta),\mathcal{D}(k))\right]\geq \delta.
	\end{align}
\end{proposition}
\begin{pf}
	For the available dataset $\mathcal{D}(k)$, the following inequalities can be obtained according to the concentration inequality:
	\begin{align}
		\mathbb{P}\left[Z(A_*,B_*)\Xi(k)\tilde{\Psi}(n(k),\delta)\Xi^{\rm T}(k)Z^{\rm T}(A_*,B_*)\succeq{\bm 0}\right]&\geq \delta.\notag
	\end{align}
	Furthermore, by recalling \eqref{eq:defGama} and 
	\begin{align}
		&Z(A_*,\!B_*)\Xi(k)\tilde{\Psi}(n(k),\delta)\Xi^{\rm T}(k)Z^{\rm T}(A_*,\!B_*)\!\succeq\!{\bm 0}\notag\\
		\Leftrightarrow&(A_*,\!B_*)\in \Gamma({\Psi}(n(k),\!\delta),\!\mathcal{D}(k)),
	\end{align}
	we obtain $\mathbb{P}\left[(A_*,B_*)\in \Gamma({\Psi}(n(k),\delta),\mathcal{D}(k))\right]\geq \delta$, which completes the proof.
\end{pf}


	%
	%

	Then, we write
	\begin{align}\label{eq:propdefAhBh}
		[\hat{A}~\hat{B}]:=X_{+}(k)\left[\begin{array}{c}
			X_{-}(k)\\ U_{-}(k)
		\end{array}\right]^{\dagger}.
	\end{align}
	To facilitate the design of ITL mechanism, the geometric property of the set-membership learning method is summarized in the following Theorem.

	\begin{theorem}\label{prop-learn}
		Consider the set $\Gamma_\delta(\mathcal{D}(k))$ parameterized in \eqref{eq:defGama}-\eqref{eq:Phi2def}. The learned set $\Gamma_\delta(\mathcal{D}(k))$ can be equivalently rewritten as a convex set in the form of  
		\begin{align}\label{eq:Gadel}
			&\Gamma_\delta(\mathcal{D}(k)\!)\!\!\\
			=&\!\Bigg\{\![A~B]\!\left|\!\left(\![A~B]\!-\![\hat{A}~\hat{B}]\!\right)\!\Phi_2(\mathcal{D}(k)\!)\!\left(\![A~B]\!-\![\hat{A}~\hat{B}]\!\right)^{\rm T}\right.\!\!\!\!\preceq\!\tilde{\Phi}_1\!\!\Bigg\},\notag
		\end{align}
		where $\Phi_2(\mathcal{D}(k))$ and $\tilde{\Phi}_1$ are parameters defined as
		\begin{align}
			&\Phi_2(\mathcal{D}(k)):=\left[\begin{array}{c}
				\!\!X_{-}(k)\!\!\\ \!\!U_{-}(k)\!\!
			\end{array}\right]\!\!\left[\begin{array}{c}
				\!\!X_{-}(k)\!\!\\ \!\!U_{-}(k)\!\!
			\end{array}\right]^{\rm T},\\
			&\tilde{\Phi}_1\!
			:=\!\Phi_1(n(k),\!\delta)\!-\!X_+(k)\!\left(\!\!I\!-\!\left[\begin{array}{c}
				\!\!X_{-}(k)\!\!\\ \!\!U_{-}(k)\!\!
			\end{array}\right]^\dagger\!\!\left[\begin{array}{c}
				\!\!X_{-}(k)\!\!\\ \!\!U_{-}(k)\!\!
			\end{array}\right]\!\right)\!X_+^{\rm T}(k).\notag
		\end{align}
	\end{theorem}
	\begin{pf}
		The set-membership learning method defined in \eqref{eq:defGama}-\eqref{eq:Phi2def} leads to a set $\Gamma({\Psi}(n(k),\delta),\mathcal{D}(k))$ in the form of
		\begin{align}
			\left\{(A,B)\!\left|Z(A,B)\Xi(k)\tilde{\Psi}(n(k))\Xi^{\rm T}(k)Z^{\rm T}(A,B)\succeq 0\right.\!\right\}.\notag
		\end{align}
		
		For $\forall (A~B)\in\Gamma_\delta(\mathcal{D}(k))$, we define a matrix $\bar{W}_-(k)$ as
		\begin{align}\label{eq:prop-learn-defW}
			\bar{W}_-(k)&:=X_{+}(k)-[A~B]\left[X_-^{\rm T}(k)~U_-^{\rm T}(k)\right]^{\rm T},
		\end{align}
		and the inequality $\bar{W}_-(k)\bar{W}^{\rm T}_-(k)\preceq \Phi_1(n(k),\delta)$ further leads to 
		\begin{align}\label{eq:prop-defWPW}
			\left(X_{+}(k)\!-\![A~B]\left[\begin{array}{c}
				\!\!X_{-}(k)\!\!\\ \!\!U_{-}(k)\!\!
			\end{array}\right]\right)\star ^{\rm T}\!\!\!
			\preceq\Phi_1(n(k),\delta).
		\end{align}
		By recalling the Moore-Penrose inverse of the matrix $[X_-^{\rm T}(k)~U_-^{\rm T}(k)]^{\rm T}$, we have
		\begin{align}
			\left[\!\!\begin{array}{c}
				X_-(k)\\U_-(k)
			\end{array}\!\!\right]^\dagger=\left[\begin{array}{c}
				\!\!X_{-}(k)\!\!\\ \!\!U_{-}(k)\!\!
			\end{array}\right]^{\rm T}\Phi_2^{-1}(\mathcal{D}(k)),\notag
		\end{align}
		and inequality \eqref{eq:prop-defWPW} can be rewritten as 
		\begin{align}
			&\left(\![A~B]\!-\![\hat{A}~\hat{B}]\!\right)\!\Phi_2(\mathcal{D}(k)\!)\!\left(\![A~B]\!-\![\hat{A}~\hat{B}]\!\right)^{\rm T}\!\!\!\\
			&+\!\!X_+(k)\!\left(I\!-\!\left[\!\!\begin{array}{c}
				X_-(k)\\U_-(k)
			\end{array}\!\!\right]^\dagger\!\!\left[\begin{array}{c}
				\!\!X_{-}(k)\!\!\\ \!\!U_{-}(k)\!\!
			\end{array}\right]\right)X_+^{\rm T}(k)\notag\\
			\preceq&\Phi_1(n(k),\delta),\notag
		\end{align}
		which completes the proof. 
	\end{pf}
	
	\begin{remark}
		Theorem \ref{prop-learn} indicates the relationship between the proposed set-membership learning method defined in \eqref{eq:defGama}-\eqref{eq:Phi2def} and the point-valued estimate \eqref{eq:propdefAhBh}.
		The estimate defined in \eqref{eq:propdefAhBh} is generally known as the least squares estimate of unknown parameters $(A_*,B_*)$.
		According to Theorem~\ref{prop-learn}, the set obtained in \eqref{eq:AtBt} is a convex set centered at the least squares estimate $[\hat{A}~\hat{B}]$, and the specific form is also influenced by the utilized data samples $[X_-^{\rm T}(k)~U_-^{\rm T}(k)]^{\rm T}$ and the matrix parameter ${\Phi}_1(n(k),\delta)$.	
	\end{remark}
	
	\subsection{Information-triggered Learning}
	In this section, we focus on the online update of the data index set $\mathcal{R}(k)$. Define
	\begin{align}\label{eq:Tk}
		T(k)&:=\left(\left[\begin{array}{c}
			\!\!X_{-}(k)\!\!\\\!\!U_{-}(k)\!\!
		\end{array}\right]\left[\begin{array}{c}
			\!\!X_{-}(k)\!\!\\\!\!U_{-}(k)\!\!
		\end{array}\right]^{\rm T}\right)^{-1}.
	\end{align}
	Then the ITL mechanism can be described by the update of the index set $\mathcal{R}(k)$ as
	\begin{align}
		&\mathcal{R}(k)\!=\!\left\{\begin{array}{ll}
			\mathcal{R}(k-1), & \text{if}~\text{\eqref{eq:ieqif2}-\eqref{eq:ieqif3} holds}, \\
			\mathcal{R}(k-1) \!\cup\!\{k\}, & \text{otherwise},
		\end{array}\right.\label{eq:etl2}
	\end{align}
	\begin{align}
		&\left|\tilde{T}(k)\right|{\rm Tr}\left[\tilde{\Phi}_1(n(k\!-\!1)\!+\!1,\delta,\hat{\mathcal{D}}(k))\right]\notag\\
		\geq& \epsilon_l^{\frac{2}{n_x}}\left|T(k-1)\right|{\rm Tr}\left[\tilde{\Phi}_1(n(k\!-\!1),\delta,\mathcal{D}(k))\right],\label{eq:ieqif2}\\
		&\Gamma(\Psi(n(k)\!+\!1,\delta),\hat{\mathcal{D}}(k))\!\subset\! \Gamma(\Psi(n(k),\delta),\mathcal{D}(k))\label{eq:ieqif3}.
	\end{align}
	where $\epsilon_l\in(0,1)$ is a constant, and $\tilde{T}(k), \hat{\mathcal{D}}(k)$ are
	\begin{align}
		\tilde{T}(k)\!&=\!\left(\left[\begin{array}{c}
			\!\!X_{-}(k)\!\!\\\!\!U_{-}(k)\!\!
		\end{array}\right]\left[\begin{array}{c}
			\!\!X_{-}(k)\!\!\\\!\!U_{-}(k)\!\!
		\end{array}\right]^{\rm T}\!\!\!+\!\!\left[\begin{array}{c}
			\!\!{\bm x}(k)\!\!\\\!\!{\bm u}(k)\!\!
		\end{array}\right]\left[\begin{array}{c}
			\!\!{\bm x}(k)\!\!\\\!\!{\bm u}(k)\!\!
		\end{array}\right]^{\rm T}\right)^{-1}\!\!\!\!,\label{eq:tildeTkp}\\
		\hat{\mathcal{D}}(k)&=\mathcal{D}(k)\cup\{{\bm x}(k),{\bm u}(k),{\bm x}(k+1)\}.
	\end{align} 

	In \eqref{eq:etl2}, an updating mechanism is proposed for the index set $\mathcal{R}(k)$, based on which the data set $\mathcal{D}(k)$ and the learned stochastic parametric uncertainty set $\Gamma({\Psi}(n(k),\delta),\mathcal{D}(k))$ can be updated accordingly.

For the ITL mechanism, the asymptotic property of the set $\Gamma({\Psi}(n(k),\delta),\mathcal{D}(k))$ is further analyzed in the following result.
	
	\begin{theorem}\label{thm:convlearn}
		Consider the system in \eqref{eq:sys} and the disturbance ${\bm w}(k)$ satisfying \eqref{eq:introdelta}. 
		If the dataset $\mathcal{D}(k)$ is updated according to the ITL mechanism \eqref{eq:etl2}, the Lebesgue measure of the set $\Gamma({\Psi}(n(k),\delta),\mathcal{D}(k))$ converges exponentially with the increase of $n(k)$ such that
		\begin{align}\label{eq:BB0}
			\mathfrak{V}\left(\Gamma({\Psi}(n(k),\delta),\mathcal{D}(k))\right)\leq \epsilon_l^{n(k)}\mathfrak{V}_0,~k>k_0
		\end{align}
		holds for some $\mathfrak{V}_0>0$ and 
		\begin{align}
			k_0&=\inf\left\{k|{\rm rank}([X_-^{\rm T}(k)~U_-^{\rm T}(k)]^{\rm T})=n_x+n_u\right\}.\notag
		\end{align}
	\end{theorem}
	\vspace{-1cm}
	\begin{pf}
		From Theorem \ref{prop-learn}, we have 
		\begin{align}
			&{\rm Tr}\Bigg[\tilde{\Phi}_1(n(k),\!\delta,\mathcal{D}(k))\notag\\
			&~~~-\left(\![A~B]\!-\![\hat{A}~\hat{B}]\!\right)\!\Phi_2(\mathcal{D}(k)\!)\!\left(\![A~B]\!-\![\hat{A}~\hat{B}]\!\right)^{\rm T}\!\!\Bigg]\geq 0.\label{eq:trace2}
		\end{align}
		To calculate the Lebesgue measure of the set $\Gamma({\Psi}(n(k),\delta),$\\$\mathcal{D}(k))$, we introduce the vectorization function  ${\rm Vec}([A~B]):$\\$~\mathbb{R}^{n_x\times(n_x+n_u)}\rightarrow\mathbb{R}^{(n_x^2+n_xn_u)\times 1}$. From \eqref{eq:defGama}-\eqref{eq:Phi2def} and the inequality in \eqref{eq:trace2}, we have	
		\begin{align}
			&{\rm Vec}\left([A~B]-[\hat{A}~\hat{B}]\right)^{\rm T}\left(\left[\!\!\begin{array}{c}
				X_{-}(k)\\ U_{-}(k)
			\end{array}\!\!\right]\left[\!\!\begin{array}{c}
				X_{-}(k)\\ U_{-}(k)
			\end{array}\!\!\right]^{\rm T}\otimes I_{n_x}\right)\notag\\
			&~~~\cdot
			{\rm Vec}\left([A~B]-[\hat{A}~\hat{B}]\right)\notag\\
			\leq&{\rm Tr}\left[\tilde{\Phi}_1(n(k),\delta,\mathcal{D}(k))\right].\label{eq:VecVec}
		\end{align}
		By defining $\tilde{\Gamma}({\Psi}(n(k),\delta),\mathcal{D}(k))$ as
		\begin{align}
			&\tilde{\Gamma}({\Psi}(n(k),\delta),\mathcal{D}(k))
			\!\notag\\
			:=&\!\left\{\![\!A~B]~\!\!\Big|\!~\text{Inequality} ~\eqref{eq:VecVec} ~\text{holds}\!\right\},
		\end{align}
		\eqref{eq:trace2} leads to the inclusion relation as
		\begin{align}
			{\Gamma}({\Psi}(n(k),\delta),\mathcal{D}(k))\subseteq\tilde{\Gamma}({\Psi}(n(k),\delta),\mathcal{D}(k)).
		\end{align}
		From \eqref{eq:VecVec}, the mapped set ${\rm Vec}(\tilde{\Gamma}({\Psi}(n(k),\delta),\mathcal{D}(k)))$ is an ellipsoid in $\mathbb{R}^{(n_x^2+n_xn_u)\times 1}$. Thus we claim that the image of the set ${\Gamma}({\Psi}(n(k),\delta),\mathcal{D}(k))$ in $\mathbb{R}^{(n_x^2+n_xn_u)\times 1}$ is enveloped by the ellipsoid defined in \eqref{eq:VecVec}. According to \cite[Section 2.1]{ellbook}, the volume of the ellipsoid ${\rm Vec}(\tilde{\Gamma}({\Psi}(n(k),\delta),\mathcal{D}(k)))$ can be calculated as
		\begin{align}
			&\mathfrak{V}\left({\rm Vec}\left(\tilde{\Gamma}({\Psi}(n(k),\delta),\mathcal{D}(k))\right)\right)\notag\\
			=&\left(\left|T(k)\right|{\rm Tr}[\tilde{\Phi}_1(n(k),\delta,\mathcal{D}(k))]\right)^{\frac{n_x}{2}}.
		\end{align}
		
		Now we are ready to analyze the convergence of the ITL mechanism designed in \eqref{eq:etl2}-\eqref{eq:tildeTkp}. In the case that $\mathcal{R}(k)=\mathcal{R}(k-1)$, the number $n(k)$, the dataset $\mathcal{D}(k)$, and the size $\mathfrak{V}({\Gamma}({\Psi}(n(k),\delta),\mathcal{D}(k)))$ do not change compared to the time instant $k-1$. Thus the following analysis concentrates on the case that $\mathcal{R}(k)=\mathcal{R}(k-1)\cup\{k\}$.
		
		In the case $\mathcal{R}(k)=\mathcal{R}(k-1)\cup\{k\}$, the following inequality can be obtained according to \eqref{eq:ieqif2}:
		\begin{align}
			&\mathfrak{V}(\tilde{\Gamma}({\Psi}(n(k),\delta),\mathcal{D}(k)))\notag\\
			<&\epsilon_l\mathfrak{V}(\tilde{\Gamma}({\Psi}(n(k-1),\delta),\mathcal{D}(k-1))).\label{eq:BGt}
		\end{align}
		Then, by writing
		\begin{align}
			\mathfrak{V}_0&=	\mathfrak{V}\left(\tilde{\Gamma}({\Psi}(n(k_0),\delta),\mathcal{D}(k_0))\right),\notag
		\end{align}
		the proof of Theorem \ref{thm:convlearn} is completed.
	\end{pf}
	
	\begin{remark}
		In the proof Theorem \ref{thm:convlearn}, the relationship between the set $\Gamma_\delta(\mathcal{D}(k))$ parameterized in \eqref{eq:defGama}-\eqref{eq:Phi2def} and an ellipsoid is characterized in \eqref{eq:VecVec}.
		The relationship facilitates the convergence analysis of the ITL mechanism in \eqref{eq:etl2} by converting the calculation of the measure of a matrix ellipsoid to the calculation of the volume of a vector ellipsoid, which is easier to calculate and analyze with the tools of ellipsoidal analysis \citep{ellbook}. Specifically, the image of the set $\Gamma({\Psi}(n(k),\delta),\mathcal{D}(k))$ in space $\mathbb{R}^{(n_x^2+n_xn_u)\times 1}$ is contained in an ellipsoid ${\rm Vec}\left(\tilde{\Gamma}(\delta,\Psi(n(k)),\mathcal{D}(k))\right)$.
		The convergence property of the learned set $\Gamma({\Psi}(n(k),\delta),\mathcal{D}(k))$ can thus indirectly obtained by analyzing the ellipsoid.
	\end{remark}

		\begin{corl}\label{corl:1}
		Let $\hat{\mathcal{I}}_{\rm QII}(k)$ be the estimate of quantitative incremental information ${\mathcal{I}}_{\rm QII}(k)$ as
		\begin{align}
			\hat{\mathcal{I}}_{\rm QII}(k)
			:=1\!-\!\frac{\mathfrak{V}\left(\tilde{\Gamma}\left(\Psi(n(k\!-\!1)\!+\!1,\delta),\mathcal{D}(k\!-\!1)\cup d_k\right)\right)}{\mathfrak{V}\big(\tilde{\Gamma}\left(\Psi(n(k\!-\!1),\delta),\mathcal{D}(k-1)\right)\big)}.\notag
		\end{align}
		Then the dataset $\mathcal{D}(k)$ satisfies
		\begin{align}
		\hat{\mathcal{I}}_{\rm QII}(r)>1-\epsilon_l, \forall r\in\mathcal{R}(k).
		\end{align} 
		\end{corl}
		\begin{pf}
			This result can be proved using \eqref{eq:BGt} and thus is omitted.
		\end{pf}

		\begin{remark}
			In Definition \ref{def:epinnovative}, the quantitative incremental information is defined as the relative reduction of the measure $\mathfrak{V}(\Gamma_\delta(\mathcal{D}(k-1)))$ caused by the addition of a data sample $d(k)$. However, the measure $\mathfrak{V}(\Gamma_\delta(\mathcal{D}(k-1)))$ is difficult to calculate directly, which poses challenges in estimating the quantitative incremental information of the data sample $d(k)$. In the proof of Theorem \ref{thm:convlearn}, an overestimate of the set $\Gamma_\delta(\mathcal{D}(k-1))$ (parameterized by $\Psi(n(k-1),\delta)$) is introduced, based on which Lemma 1 provides a method for estimating the quantitative incremental information, namely, $\hat{\mathcal{I}}_{\rm QII}(k)$. Corollary 1 demonstrates that the estimated quantitative incremental information $\hat{\mathcal{I}}_{\rm QII}(k)$ of all selected data samples exceeds a threshold $1-\epsilon_l$. In other words, Corollary 1 indicates the characteristics of the designed event-triggering mechanism in \eqref{eq:etl2} as $\{k\}\in\mathcal{R}(k)\Leftrightarrow\hat{\mathcal{I}}_{\rm QII}(k)>1-\epsilon_l$.
		\end{remark}

	Next, we briefly show how to find a stabilizing controller with the proposed ITL. Before continuing, we first recall the definition of informativity-based stabilizing controller (ISC) (which is a generalized version of Definition 3 in \cite{van2020noisy}).

	\begin{define}[ISC]\label{def:informative_controller}
		Let $\mathcal{D}(k)$ be a set of data samples of system \eqref{eq:sys}, and $\Gamma({\Psi}(n(k),\delta),\mathcal{D}(k))$ be a set of unknown system parameters defined as \eqref{eq:defGamma}. 
		Then, a controller  $\mathfrak{C}({\bm x}): \mathbb{R}^{n_x}\rightarrow\mathbb{R}^{n_u}$ is called an ISC if the controller ${\bm u}=\mathfrak{C}({\bm x})$ stabilizes all systems in the set $\Gamma({\Psi}(n(k),\delta),\mathcal{D}(k))$.
	\end{define}

	Using the above definition, a sufficient condition for the existence of an ISC is shown in Proposition \ref{lem:sufcon}.
	
	\begin{proposition}\label{lem:sufcon}
		An ISC $K(k)$ can be obtained using dataset $\mathcal{D}(k)$ if there exist matrices $P\in\mathbb{R}^{n_x\times n_x},~P=P^{\rm T}\succ0$, $L\in\mathbb{R}^{n_u\times n_x}$, and scalars $\xi\geq 0,~\rho>0$ satisfying 
		\begin{align}\label{eq:sufcon1}
			\left[\begin{array}{cccc}
				P-\rho I_{n_x}& {\bm 0} & {\bm 0} & {\bm 0}\\
				{\bm 0} &-P&-L^{\rm T}&{\bm 0}\\
				{\bm 0}&-L&{\bm 0}&L\\
				{\bm 0}&{\bm 0}&L^{\rm T}&P
			\end{array}\right]-\xi\check{\Xi}(k)\Psi(n(k))\check{\Xi}^{\rm T}(k)\succeq 0,
		\end{align}
		where
		\begin{equation}
			\check{\Xi}(k):=\left[\begin{array}{c}
				\Xi(k)\\{\bm 0}_{n_x\times(n(k)+n_x)}
			\end{array}\right].\notag
		\end{equation}
		
		Moreover, if $P$ and $L$ satisfy \eqref{eq:sufcon1}, then $K(k):=LP^{-1}$ is a stabilizing feedback gain for $(A,B)\in\Gamma({\Psi}(n(k),\delta),\mathcal{D}(k))$.
	\end{proposition}
	
	\begin{pf}
		This result can be proved following a similar line of arguments to the proof of Theorem 14 in \cite{van2020noisy} with the help of the matrix S-lemma \citep{van2023quadratic}, and thus the proof is omitted.
	\end{pf}

	Using Proposition \ref{lem:sufcon}, Algorithm \ref{alg:updating} provides the implementation details of the proposed ITL method. Specifically, when the available dataset $\mathcal{D}(k)$ does not exhibit persistent excitation, all data samples are included in the dataset (Lines \ref{line:begin}-\ref{line:L1}). Once the persistent excitation condition ${\rm rank}([X_-^{\rm T}(k)~U_-^{\rm T}(k)]^{\rm T})=n_x+n_u$ is satisfied, an ITL condition is evaluated to prevent the use of excessive redundant data (Lines \ref{line:L2}-\ref{line:L3}). If the new sample $d(k)$ does not provide sufficient information, the dataset is left unchanged (Lines \ref{line:L4}). Otherwise, the new sample is included in the dataset $\mathcal{D}(k)$, and the index set $\mathcal{R}(k)$ is updated accordingly (Lines \ref{line:L6}-\ref{line:L7}).
		\begin{algorithm}
		\caption{Online data updating}
		\label{alg:updating}
		\begin{algorithmic}[1]
		\STATE \textbf{Input:} learning hyperparameter $\epsilon_l$;
		\STATE Receive samples $d(0)={\bm x}(0),{\bm u}(0),{\bm x}(1)$;\label{line:begin}
		\STATE Initialization $\mathcal{R}(0)\!=\!\{0\},~\mathcal{D}(0)\!=\!\{d(0)\},~n(0)\!=\!1$;
		\FOR{$k=1,2,3,\ldots$}
			\STATE Receive a sample $d(k)={\bm x}(k),{\bm u}(k),{\bm x}(k+1)$;
			\IF{${\rm rank}([X_-^{\rm T}(k)~U_-^{\rm T}(k)]^{\rm T})<n_x+n_u$}
				\STATE $\mathcal{R}(k)=\mathcal{R}(k-1)\cup\{k\}$;
				\STATE Update $\mathcal{D}(k)$ and $n(k)$ accordingly;\label{line:L1}
			\ELSE
				\STATE Calculate $T(k-1)$, $\tilde{T}(k)$ according to \eqref{eq:Tk} and \eqref{eq:tildeTkp}, respectively;\label{line:L2}
				\STATE Calculate ${\rm Tr}\left[\tilde{\Phi}_1(n(k\!-\!1)\!+\!1,\delta,\hat{\mathcal{D}}(k))\right]$ and ${\rm Tr}\left[\tilde{\Phi}_1(n(k\!-\!1),\delta,\mathcal{D}(k))\right]$;
				\IF{Inequalities \eqref{eq:ieqif2}-\eqref{eq:ieqif3} holds}\label{line:L3}
					\STATE $\mathcal{R}(k), \mathcal{D}(k), n(k), K(k)$ remain unchanged	\label{line:L4}		
					\ELSE
					\STATE $\mathcal{R}(k)=\mathcal{R}(k-1)\cup\{k\}$;\label{line:L6}
					\STATE Update $\mathcal{D}(k)$ and $n(k)$ accordingly;
					\STATE Resolve the LMI \eqref{eq:sufcon1} to update $K(k)$;\label{line:L7}
				\ENDIF
			\ENDIF
		\ENDFOR
		\end{algorithmic}
		\end{algorithm}

	\subsection{Discussions on Special Concentration Inequalities}\label{sec:conineq}
	Different parameterizations of the concentration inequality in \eqref{eq:introdelta} can be obtained if different prior knowledge of the disturbance is available. In this subsection, we discuss two important examples, including deterministic disturbances with known upper bounds and stochastic disturbances with known covariance; additional examples can be referred to \cite{van2023quadratic,brailovskaya2024universality,brunzema2024neural}.
	
	{\textbf Case I: Deterministic disturbances with known upper bounds.}
	For bounded disturbance ${\bm w}(k){\bm w}^{\rm T}(k)\preceq\bar{\phi}$ with parameters set as $\delta=1,~\Phi_1(n(k),\delta)=n(k)\bar{\phi}$,
	an equation in the form of \eqref{eq:introdelta} can be obtained as $\mathbb{P}\left[W_-(k)W_-^{\rm T}(k)\preceq n(k)\bar{\phi}\right]=1.$
	This case is practical since disturbance signals in engineering applications are usually bounded. A conservative upper bound $\bar{\phi}$ is sufficient to obtain the parameter $\Phi_1(n(k),\delta)$ in \eqref{eq:introdelta}.

	{\textbf Case II: Stochastic disturbances with covariance information.}
	Let ${\bm \sigma}_w$ be the covariance matrix of the stochastic disturbance ${\bm w}(k)$. Then for the random matrix
	\begin{align}
		W_{_-}(k)W_{_-}^{\rm T}(k)=\sum\limits_{i\in\mathcal{R}(k)}{\bm w}(i){\bm w}^{\rm T}(i)\succeq0,\notag
	\end{align}
	the expectation of which can be denoted as
	\begin{align}
		\mathbb{E}[W_{_-}(k)W_{_-}^{\rm T}(k)]=n(k){\bm \sigma}_{\bm w}.\notag
	\end{align}
	
	Before discussing the parameterization of the concentration inequality for this case, we introduce an instrumental lemma as follows.
	\begin{lemma}[\cite{ahlswede2002strong}]\label{lem:marineq}
		Let $\Phi\succ0$ be a matrix parameter, and let $X$ be a random matrix such that $X\succ 0$ almost surely. Then the following inequality holds:
		\begin{align}\label{eq:makeq}
			\mathbb{P}\left[X\npreceq\Phi\right]\leq {\rm Tr}(\mathbb{E}[X]\Phi^{-1}).
		\end{align}
	\end{lemma}

	According to Lemma \ref{lem:marineq}, an inequality can be obtained as
	\begin{align}
		&\mathbb{P}(\Phi(n(k))-W_{_-}(k)W_{_-}^{\rm T}(k)\succeq0)\\
		\geq& 1-{\rm Tr}(\mathbb{E}[W_{_-}(k)W_{_-}^{\rm T}(k)]\Phi^{-1}(n(k))),\notag
	\end{align}
	and a concentration inequality for Case II can be obtained in the following lemma.

	\begin{lemma}\label{lem:chosPhi}
		For $\delta\in(0,1)$ and a random variable ${\bm w}(k)$ with covariance matrix ${\bm \sigma}_{\bm w}$, equation \eqref{eq:introdelta} holds if 
		\begin{align}\label{eq:Phicov}
			\Phi_1(n(k),\delta)=\frac{n_xn(k)}{1-\delta}{\bm \sigma}_{\bm w}.
		\end{align}
	\end{lemma}
	\begin{pf}
		According to the definition of covariance matrix, the item $W_-(k)W_-^{\rm T}(k)$ can be seen as a random matrix with an expectation being $\mathbb{E}[W_-(k)W_-^{\rm T}(k)]=n(k){\bm \sigma}_{\bm w}$. 
		For the parameter $\Phi_1(n(k),\delta)$ provided in \eqref{eq:Phicov}, the following equations hold:
		\begin{align}
			&\mathbb{P}(\Phi_1(n(k),\delta)-W_-(k)W_-^{\rm T}(k)\succeq 0)\\
			=&\mathbb{P}(W_-(k)W_-^{\rm T}(k)\preceq\Phi_1(n(k),\delta))\\
			\geq&1-{\rm Tr}(\mathbb{E}[W_-(k)W_-^{\rm T}(k)]\Phi_1^{-1}(n(k),\delta))=\delta,\notag
		\end{align}
		which completes the proof. 
	\end{pf}
	\begin{remark}
		In engineering applications, the covariance of the disturbance can be estimated from data samples even if the system parameters are unknown.
		For instance, \cite{PELCKMANS2005100} provided a model-free estimation approach for the disturbance variance. 		
		On the other hand, although an accurate estimate of the covariance ${\bm \sigma}_w$ may be unavailable in practice, it can be replaced by its upper bound $\bar{\bm \sigma}_w$ and Lemma \ref{lem:chosPhi} still holds. 
		Using an upper bound of the covariance $\bar{\bm \sigma}_w$, the parameter  $\Phi_1(n(k),\delta)$ can be selected as $\frac{n_xn(k)}{1-\delta}\bar{\bm \sigma}_{\bm w}$ 
		to ensure  inequality \eqref{eq:introdelta}, which can be proved similarly to Lemma \ref{lem:chosPhi}.
	\end{remark}

	\subsection{Applications to LPC Design}\label{sec:imp}	
	In this section, we show how the proposed ITL mechanism can be utilized to design an LPC.		
	Compared with standard predictive controllers, the proposed mechanisms lead to intermittent updates of $\mathcal{D}(k)$, $K(k)$, and $\Gamma(\Psi(n(k),\delta), \mathcal{D}(k))$, which adds challenges to closed-loop performance analysis.
	To address this issue, this section focuses on how the uncertainties induced by the ITL can be suitably dealt with to ensure the closed-loop stability of the predictive controller.

	Similar to \citep{bayer2016tube,mcallister2022stochastic}, we parameterize the controller as
	\begin{equation}\label{eq:contparame}
		\mu_c({\bm x}(k),{\bm v}(k)):={\bm v}(k)+K(k){\bm x}(k).
	\end{equation}
	Here, the feedback gain $K(k)$ is obtained from \eqref{eq:sufcon1}, and ${\bm v}(k)$ is the decision variable in predictive controller.

	The stage cost function is defined as 
	\begin{align}
		l({\bm x}, {\bm u}):={\bm x}^{\rm T}Q{\bm x}+{\bm u}^{\rm T}R{\bm u},\notag
	\end{align}
	where $Q$ and $R$ are positive definite and symmetric matrices.{ Let $\delta_M>0$ be a positive constant, then the terminal cost function is designed as
		\begin{align}
			V_f({\bm x};P_f(k)):=&{\bm x}^{\rm T}P_f(k){\bm x},\\
			P_f(k)=&Q+K^{\rm T}(k)RK(k)+\delta_MI.\label{eq:Pf}
		\end{align}

		For a series of disturbances ${\bm W}(k;N)$, let $\mathbb{W}_{\delta_1}$ be a compact set satisfying $\mathbb{P}({\bm W}(k;N)\in\mathbb{W}_{\delta_1})\geq \delta_1$.	

		Let $L_f$ be a local Lipschitz constant of the terminal cost function $V_f(\cdot)$ with the domain being a compact set $\mathbb{W}_{\delta_1}$ that contains the origin. Moreover, a terminal set can be designed in the form of $\mathbb{X}_f(k):=\{{\bm x}|{\bm x}^{\rm T}P_f(k){\bm x}\leq \theta_f\}$ (similar to \cite{doi:10.1137/18M1176671}), with $\theta_f$ obtained from
		\begin{align}
			\theta_f=&\arg\min\limits_{\theta} \theta\label{eq:defthetaf}\\
			\text{s.t.}	&~~~\theta\geq {\bm x}^{\rm T}\delta_M {\bm x}+L_f\max\limits_{{\bm w}\in\mathbb{W}_{\delta_1}}\|{\bm w}\|,~{\bm x}\in\mathbb{X}.\notag
		\end{align}

	}
	
	Let $N$ be the length of the prediction horizon, and write a series of control inputs as 
	\begin{align}
		{\bm \mu_c}(k;N):=&\{\mu_c({\bm x}(k),{\bm v}(k)),\mu_c(\bar{\bm x}(k+1),{\bm v}(k+1)),\\
		&~~~\ldots,\mu_c(\bar{\bm x}(k+N),{\bm v}(k+N))\},\notag
	\end{align}
	where $\bar{\bm x}(k+i), i\in\{1,2,\ldots,N\}$ is the predicted state defined as 
	\begin{align}
		\bar{\bm x}(k+i):=\Phi_p(i;{\bm x}(k),{\bm \mu}_c(k;N),\hat{\bm W}(k;N)).\notag
	\end{align}
	Here, $\Phi_p(i;({\bm x}(k),{\bm \mu}_c(k;N),\hat{\bm W}(k;N)))$ is the solution of the prediction model with $A,B$ as
	\begin{align}\label{eq:predict}
		&\bar{\bm x}(k+i+1)\\
		=&{A}\bar{\bm x}(k+i)+{B}\mu_c(\bar{\bm x}(k+i),{\bm v}(k+i))+\hat{\bm w}(k+i),\notag
	\end{align}
	where the initial state is $\bar{\bm x}(k)={\bm x}(k)$, $\hat{\bm w}(k+i)\in\mathbb{W}_{\delta_1}, i\in\{0,1,\ldots,N-1\}$, and $\hat{\bm W}(k;N)$ is a series of disturbances as
	\begin{align}\label{eq:defhatW}
		\hat{\bm W}(k;N)\!:=\!\{\hat{\bm w}(k),\hat{\bm w}(k+1),\ldots,\hat{\bm w}(k\!+\!N\!-\!1)\}.
	\end{align}
	The disturbances used in the prediction model \eqref{eq:predict} are marked by ``~$\hat{}$~'' to distinguish them from the unknown disturbance in system \eqref{eq:sys}.
	

	Utilizing the variables defined above, the cost function can be defined as
	\begin{align}\label{eq:costJ}
		&J_N({\bm x}(k),{\bm \mu}_c(k;\!N),\!\hat{\bm W}(k;N);\!{A},\!{B})\\
		:=&\sum\limits_{i=0}^{N-1}l(\bar{\bm x}(k\!+\!i),\!\mu_c(\bar{\bm x}(k\!+\!i),\!{\bm v}(k\!+\!i)\!)\!)\!+\!V_f(\bar{\bm x}(N)\!)\notag.
	\end{align}

Using aforementioned notations, the optimal control sequence at time instant $k$ can be defined as 
\begin{align}
	\min\limits_{V(k)}&\max\limits_{\hat{W}(k),A,B}J_N({\bm x}(k),{\bm \mu}_c(k;N),\hat{\bm W}(k;N);{A},{B})\label{eq:MPC1}\\
	{\rm {s.t.}}
		&~~~~ \bar{\bm x}(k)={\bm x}(k),~\hat{\bm W}(k;N)\in\mathbb{W}_{\delta_1},\notag\\
		&~~~~({A},{B})\in{\Gamma}({\Psi}({n}(k),\delta),{\mathcal{D}}(k)),\label{eq:consAB}\\
		&~~~~\bar{\bm x}(k\!+\!i)=A\bar{\bm x}(k\!+\!i\!-\!1)+\hat{\bm w}(k\!+\!i\!-\!1)\label{eq:Xpred}\\
		&~~~~~~~~~~~~~~~~~~~+B\mu_c(\bar{\bm x}(k\!+\!i\!-\!1),{\bm v}(k\!+\!i\!-\!1)),\notag\\
		&~~~~\mu_c(\bar{\bm x}(k\!+\!i),{\bm v}(k\!+\!i))=K(k)\bar{\bm x}(k\!+\!i)+{\bm v}(k\!+\!i),\notag\\
		&~~~~\mu_c(\bar{\bm x}(k+i),{\bm v}(k+i))\in\mathbb{U},~\bar{\bm x}(k+i)\in\mathbb{X}.\notag
\end{align}

In the optimization problem \eqref{eq:MPC1}, $\mathbb{X}$ and $\mathbb{U}$ represent the state and input constraints, respectively. These constraints are determined by operational limits, physical restrictions, and cost requirements in practice. 
\newline
Moreover, it is important to note that the input and state constraints must hold for all parameters satisfying \eqref{eq:consAB}. They must also hold for all disturbances within $\hat{\bm W}(k;N)\in\mathbb{W}_{\delta_1}$. This is because the controller is designed to be robust against all dynamic uncertainties in the set $\Gamma({\Psi}(n(k),\delta),\mathcal{D}(k))$ and disturbances in $\mathbb{W}_{\delta_1}$. The problem can be solved using reformulation and scenario-based approaches, which is summarized in Appendix.~\ref{App:1-solving}.

	 Write the optimal cost function as $V_N^*({\bm x}(k))$ and the optimal sequence for ${\bm v}(k+i)$ as ${\bm v}^*(k+i),i\in\{0,\ldots,N-1\}$. Then the optimal control sequence obtained at the time instant $k$ is denoted as ${\bm \mu}_c^*(k;N)$.


	\begin{lemma}\label{lem:Pfinv}
		Consider the system in \eqref{eq:sys} with dataset $\mathcal{D}(k)$ obtained according to the ITL mechanism \eqref{eq:etl2} and the terminal cost function $V_f(\cdot;P_f(k))$ obtained as \eqref{eq:Pf}. The inequality
		\begin{align}\label{eq:cond1}
			&V_f(A{\bm x}(k)+BK(k){\bm x}(k);P_f(k+1))\notag\\
			\leq& V_f({\bm x}(k);P_f(k))-l({\bm x}(k),{\bm u}(k))
		\end{align}
		holds for ${{\bm x}\in\mathbb{X}}$, $(A,B)\in\Gamma({\Psi}(n(k),\delta),\mathcal{D}(k))$ if there exists $\xi>0$ such that 
		\begin{align}\label{eq:Deltaxi}
			&\left[\begin{array}{cc}
				P_f^{-1}(k+1)&{\bm 0}\\
				{\bm 0}&-\left[\begin{array}{c}
					I\\K(k)
				\end{array}\right]\frac{1}{\delta_M}I[I~K^{\rm T}(k)]
			\end{array}\right]\notag\\
			&~-\xi\Xi(k)\tilde{\Psi}(n(k))\Xi^{\rm T}(k)\succeq 0
		\end{align}
		holds for $\delta_M>0$ with the matrix $K(\cdot)$ obtained according to the LMI \eqref{eq:sufcon1}.
	\end{lemma}
	
	\begin{pf}
		For $(A,B)\in\Gamma({\Psi}(n(k),\delta),\mathcal{D}(k))$, the following inequality is satisfied according to \eqref{eq:defGamma}
		\begin{align}
			Z(A,B)\Xi(k)\tilde{\Psi}(n(k))\Xi^{\rm T}(k)Z^{\rm T}(A,B)\succeq 0.
		\end{align}
		Thus, the following inequality can be obtained from \eqref{eq:Deltaxi} and matrix S-lemma \citep{van2023quadratic}:
		\begin{align}\label{eq:Pfdelta}
			P_f^{-1}(k\!+\!1)\!-\!\left[A\!+\!BK(k)\right]\frac{I}{\delta_M}\left[A\!+\!BK(k)\right]^{\rm T}&\succeq\! 0.
		\end{align}
		
		
		In view of that matrix $P_f(k)$ is a symmetric and positive definite matrix according to \eqref{eq:Pf}, we denote 
		\begin{align*}
			P_f(k)&=S(k)\Lambda_f(k)S^{-1}(k),\\
			P_f^{-1}(k)&=S(k)\Lambda_f^{-1}(k)S^{-1}(k),
		\end{align*}
		where $S(k)$ is invertible matrix and $\Lambda_f(k)$ is a diagonal matrix.
		By left and right multiplying $P_f(k)$ to matrix 
		\begin{align}
			P_l(k)\!:=\!P_f^{-1}(k)\!-\!P_f^{-1}(k)[Q\!+\!K^{\rm T}(k)R(k)K(k)]P_f^{-1}(k),\notag
		\end{align}
		the following identity can be obtained
		\begin{align}
			&P_{f}(k)P_l(k)P_f(k)\label{eq:dMuse}\\
			=&P_f(k)-[Q+K^{\rm T}(k)R(k)K(k)]=\delta_MI\notag.
		\end{align}
		Then matrix $P_l(k)$ can be equivalently written as
		\begin{align}
			P_l(k)=S(k)\delta_M{\Lambda_f^{-2}(k)}S^{-1}(k),
		\end{align}
		based on which we have
		\begin{align}\label{eq:dpp}
			\frac{I}{\delta_M}&=P_f^{-1}(k)P_l^{-1}(k)P_f^{-1}(k).
		\end{align}
		
		
		With \eqref{eq:dpp}, Inequality \eqref{eq:Pfdelta} leads to 
		\begin{align}
			&P_f^{-1}(k+1)\notag\\
			\succeq&[(A+BK(k))P_f^{-1}(k)]P_l^{-1}(k)[(A+BK(k))P_f^{-1}(k)]^{\rm T}.\notag
		\end{align}
		Then, by noting that $P_l(k)$ is a positive definite matrix, the following inequality can be obtained according to Schur complement argument,
		\begin{align}
			\left[\begin{array}{cc}
				\!\!P_l(k)&\left[(A+BK(k))P_f^{-1}(k)\right]^{\rm T}\!\!\\
				\!\!\left[(A+BK(k))P_f^{-1}(k)\right]&P_f^{-1}(k+1)\!\!
			\end{array}\right]\succeq0,\notag
		\end{align}
		which further leads to 
		\begin{align}
			&P_l(k)\label{eq:Plab}\\
			\succeq&[(A\!+\!BK(k))P_f^{-1}(k)]^{\rm T}P_f(k\!+\!1)[(A\!+\!BK(k))P_f^{-1}(k)].\notag
		\end{align}
		As a result, the proof of the claim \eqref{eq:cond1} is completed by left and right multiplying $P_f(k)$ to \eqref{eq:Plab}.
	\end{pf}

The parameter $\delta_M$ is a user-defined parameter introduced in \eqref{eq:Pf} and utilized in \eqref{eq:Deltaxi}. According to Lemma \ref{lem:Pfinv}, inequality \eqref{eq:Deltaxi} provides a condition for selecting the parameter $\delta_M$. After introducing the generalized eigenvalues, the condition for selecting $\delta_M$ is presented in a proposition.

\begin{define}[Generalized eigenvalue]
	Let $P_1,P_2\in\mathbb{R}^{n\times n}$ be two real matrices. The roots of the equation ${\rm det}(P_1-\lambda P_2)=0$ are called the generalized eigenvalues of the matrix pencil $(P_1,P_2)$. The smallest generalized eigenvalue of $(P_1,P_2)$ is denoted as $\lambda_{\rm min}(P_1,P_2)$. 
\end{define}

\begin{proposition}\label{prop:lowbounddelta}
To ensure the inequality \eqref{eq:Deltaxi}, the parameter $\delta_M$ introduced in \eqref{eq:Pf} has a lower bound as 
\begin{align}
	\delta_M \geq \frac{1}{{\hat{\lambda}_{\rm min} \check{\lambda}_{\rm min}}},
\end{align} 
where $\hat{\lambda}_{\rm min}$ and $\check{\lambda}_{\rm min}$ are the smallest generalized eigenvalues of the matrix pencils $(P_f^{-1}\!(k),\Phi_1(n(k),\delta)\!-\!X_+(k)X_+^{\rm T}(k))$ and $\left(\left[\!\!\begin{array}{c}
	X_-(k)\\U_-(k)
\end{array}\!\!\right]\left[X_-(k)^{\rm T}~U_-(k)^{\rm T}\right],\right.$\\$\left.\left[\!\!\begin{array}{c}
	I\\K(k)
\end{array}\!\!\right][I~K^{\rm T}(k)]\!\!\right)$, respectively.
\end{proposition}
\begin{pf}
The proof can be completed using inequality \eqref{eq:Deltaxi}, the Schur complement, and the properties of generalized eigenvalues. The details are provided in Appendix.~\ref{app:prop1}.
\end{pf}


	%
	Note that the condition in \eqref{eq:Deltaxi} is a data-based LMI with the only unknown parameter being the scalar $\xi$, which can be easily tested. In addition, this condition only needs to be evaluated during an event instant when a new data sample is incorporated into $\mathcal{D}(k)$, since the parameters remain constant during a non-event instant.

	\begin{lemma}\label{lem:5}
		Consider ${\bm x}\in\mathbb{X}_f(k)$ and $\theta_f$ designed as \eqref{eq:defthetaf}. If \eqref{eq:cond1} holds for $A~B$, we have
		\begin{align}\label{eq:babbx}
			&\forall[A~B]\in\Gamma_\delta(\mathcal{D}(k)),~\exists {\bm u}=K(k){\bm x}(k):\notag\\
			&{A}{\bm x}+{B}{\bm u}+{\bm w}\in\mathbb{X}_f(k+1),~{\bm w}\in\mathbb{W}_{\delta_1}.
		\end{align}
	\end{lemma}
	
	\begin{pf}
		This lemma can be verified using \eqref{eq:defthetaf} and thus the proof is omitted.
	\end{pf}

The recursive feasibility of the proposed controller is discussed in the following proposition.

\begin{proposition}\label{prop:feas}
Consider the system in \eqref{eq:sys} with dataset $\mathcal{D}(k)$ obtained according to the ITL mechanism \eqref{eq:etl2} and the terminal cost function $V_f(\cdot;P_f(k))$ obtained as \eqref{eq:Pf}. If the conditions in inequality \eqref{eq:Deltaxi} hold, then the MPC problem \eqref{eq:MPC1} is recursively feasible.
\end{proposition}
\begin{pf}
The proof can be performed by demonstrating that the optimal solution at time $k$ is a feasible solution at time $k+1$, which can be guaranteed by the designed ITL mechanism. The detailed proof is provided in Appendix.~\ref{app:feas}.
\end{pf}

	For a feedback control gain $K(k)$ such that inequalities \eqref{eq:cond1} and \eqref{eq:babbx} hold, we define $\delta_2$ as
	\begin{align}\label{eq:defdelta}
		&~\delta_2
		:=\max\limits_{{\bm w},\bar{A},\bar{B}} V_f(\bar{A}{\bm x}+\bar{B}{\bm u}+{\bm w};P_f(k+1))\notag\\
		&~~~~~~~~~~~~~~~~~~~~-V_f({\bm x};P_f(k))+l({\bm x},K(k){\bm x})\\
		&{\rm s.t.} ~{\bm w}\in\mathbb{W}_{\delta_1},~~{\bm x}\in\mathbb{X},~~(\bar{A},\bar{B})\in\Gamma({\Psi}(n(k),\delta),\mathcal{D}(k)).\notag
	\end{align}


	
	In this work, the cost function \eqref{eq:costJ} is designed as a quadratic and positive definite function, thus we have
	\begin{align}
		l({\bm x},{\bm u})\geq c_1 |{\bm x}|^2,~
		V_N^*({\bm x})\leq c_2|{\bm x}|^2,~\forall {\bm x}\in\mathbb{X},\notag
	\end{align}
	with $c_1,c_2>0$ being positive constants. Now we are ready to analyze the stability of the closed-loop system by applying the designed learning-based predictive controller with the proposed {information-triggered} learning mechanism.
	
	\begin{proposition}\label{thm:stampc}
		Let $\gamma:=1-\frac{c_1}{c_2}, c:=\frac{\delta_2+\epsilon}{1-\gamma}$ with $\delta_2$ defined in \eqref{eq:defdelta} and $\epsilon>0$. For system \eqref{eq:sys} with the dataset obtained by the proposed {information-triggered} learning mechanisms such that \eqref{eq:Deltaxi} holds, the LPC \eqref{eq:MPC1} is an ISC and satisfies:
		\begin{itemize}
			
			\item the set ${\rm lev}_c(V_N^*):=\{{\bm x}|V_N^*({\bm x})\leq c\}$ is a positive invariant set with probability $\delta\delta_1$, i.e., for ${\bm x}\in{\rm lev}_c(V_N^*)$, we have
			\begin{align}\label{eq:mpcends1}
				\mathbb{P}\left(A_*{\bm x}+B_*\mu_c^*({\bm x},{\bm v}^*)+{\bm w}\in{\rm lev}_c(V_N^*)\right)\geq\delta\delta_1;
			\end{align}
			\item if ${\bm x}\in\mathbb{X}, V_N^*({\bm x})>c$, the state is steered closer to the positive invariant set ${\rm lev}_c(V_N^*)$ with a step larger than $\epsilon$ with probability $\delta\delta_1$, i.e., for $d:=V_N^*({\bm x})>c$:
			\begin{align}\label{eq:mpcends2}
				\mathbb{P}\left(V_N^*(A_*{\bm x}+B_*\mu_c^*({\bm x},{\bm v}^*)\!+\!{\bm w})\leq d\!-\!\epsilon\right)\!\geq\!\delta\delta_1.
			\end{align}
		\end{itemize}
	\end{proposition}
	
	\begin{pf}
		Given that \eqref{eq:Deltaxi} holds, inequality \eqref{eq:cond1}, and \eqref{eq:babbx} can be obtained according to Lemmas \ref{lem:Pfinv}-\ref{lem:5}, respectively. Then, the proof of this proposition can be performed {by taking $V_N^*({\bm x}(k))$ as a Lyapunov function following a similar line of arguments to Section 3.4 in \cite{MPCJB}}.
	\end{pf}

	\section{Numerical Examples}\label{sec:nbex}
	In this section, numerical results are shown to verify the proposed results and to compare with other methods.
	In what follows, the results in Theorems \ref{prop-learn} and \ref{thm:convlearn} are first validated through extensive numerical simulations. Then, comparative simulations are performed to compare the performance of the proposed ITL-based predictive control approach with DeePC \citep{coulson2019data} and robust reinforcement learning (RRL) \citep{umenberger2019robust}. 
	

	\subsection{Illustration of Theorems \ref{prop-learn}}
\begin{figure}[tbp]
	\centering
	\subfloat[$\bar{\sigma}=0.01$]{
		\includegraphics[width=0.88\linewidth]{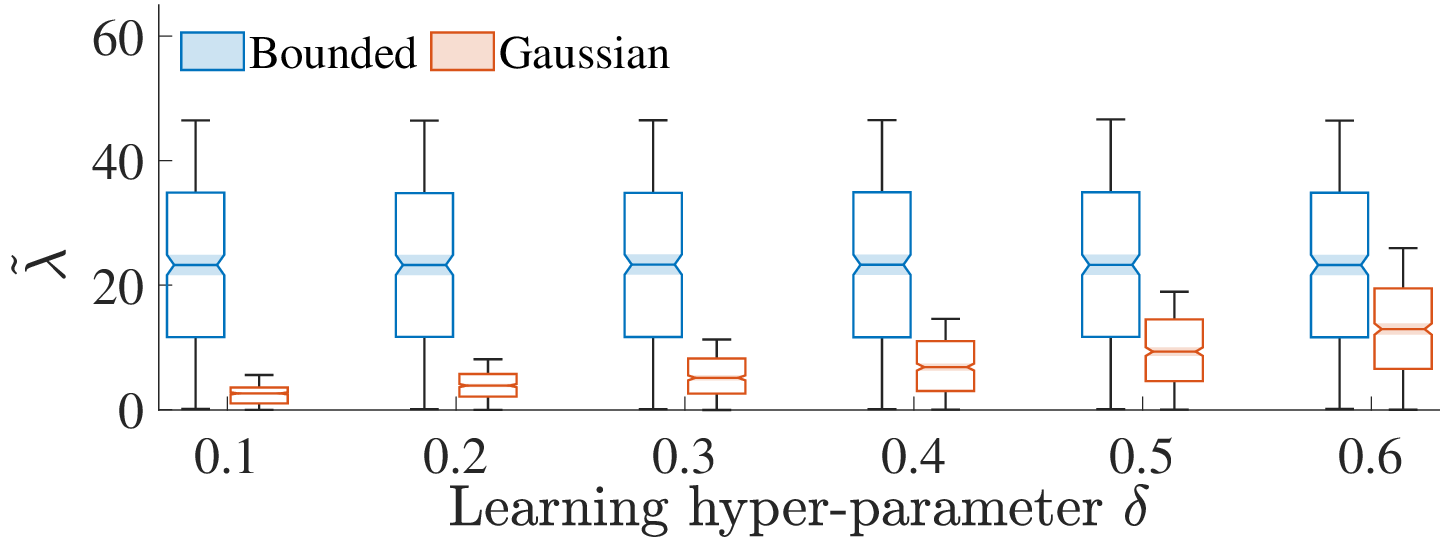}
	}\\
	\subfloat[$\bar{\sigma}=0.3$]{
		\includegraphics[width=0.88\linewidth]{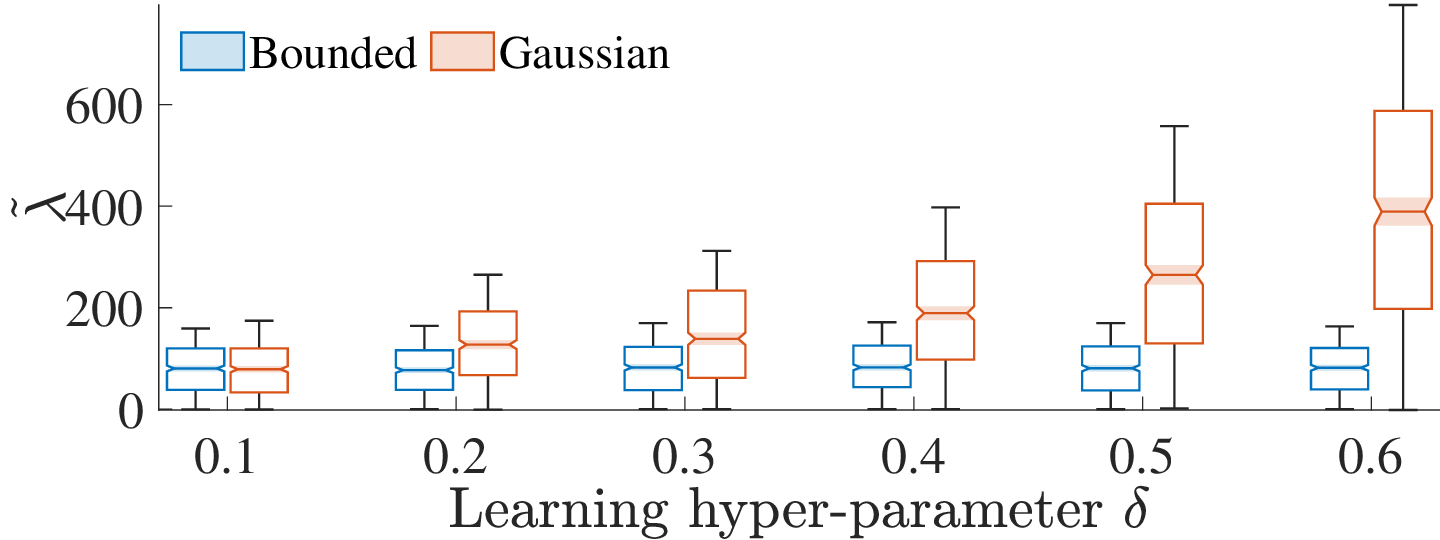}
	}
	\caption{Statistical results of $6,000$ independent numerical experiments. The red box-plots denote results for bounded disturbances with bound being $\bar{\phi}=n_x\sqrt{\bar{\sigma}}$, and the blue box-plots represent results for Gaussian disturbances with covariance matrix being $\sigma_{\bm w}=\bar{\sigma}I_{n_x}$. }
	\label{fig:thm1}
\end{figure}

In this subsection, $6,000$ third-order linear systems are generated randomly with $\lambda_i(A)\in[-3,3],~i\in\{1,2,3\}$. 
For each linear system, a trajectory is generated with random initial states and inputs. 
We recall that, if $(A_*,B_*)\in\Gamma({\Psi}(n(k),\delta),\mathcal{D}(k))$, then the following inequality holds:
\begin{align}
	&Z(A_*,B_*){\Psi}(n(k),\delta)Z^{\rm T}(A_*,B_*)\succeq 0\\
	\Rightarrow&\lambda_{n_x}(Z(A_*,B_*){\Psi}(n(k),\delta)Z^{\rm T}(A_*,B_*))>0,
\end{align}
where $\lambda_{n_x}(\cdot)$ denotes the minimum eigenvalue of a matrix.
To validate this relationship, box plots \citep{9968068} of the minimum eigenvalues are shown in Fig.~\ref{fig:thm1}, which indicates that the minimum eigenvalue of the matrix $Z(A_*,B_*){\Psi}(n(k),\delta)Z^{\rm T}(A_*,B_*)$ in each experiment is larger than zero. 
This result verifies the property of the proposed set-membership learning method (Theorem \ref{prop-learn}) because all the estimates satisfy  $(A_*,B_*)\in\Gamma({\Psi}(n(k),\delta),\mathcal{D}(k))$.

\subsection{Comparative Simulations}
In this section, the proposed ITL protocol and learning-based predictive control method are compared with other learning-based controllers, including the DeePC proposed in \cite{coulson2019data} and RRL proposed in \cite{umenberger2019robust}. To do this, consider a linear time-invariant {controllable} system of the form \eqref{eq:sys} with $A_*$ and $B_*$ being
\begin{align}
	A_*\!\!=\!\!&\left[\!\!\begin{array}{ccc}
		0.850 &-0.038& -0.038\\0.735& 0.815& 1.594\\ -0.664 &0.697&-0.064
		\end{array}\!\!\right],~
	B_*\!\!=\!\!\left[\!\!\begin{array}{cc}
		1.431& 0.705\\1.620& -1.129\\0.913 &0.369
	\end{array}\!\!\right],\label{eq:sysdat}
\end{align}
 which is the example used in \cite{van2020noisy}. This system is unstable since the eigenvalues of matrix $A_*$ are $\{-0.7664, 0.8536, 1.5138\}$. The input signal ${\bm u}$ is designed to be a white disturbance before an ISC is obtained, and the state is initialized as ${\bm x}(1)=[0~0~0]^{\rm T}$. After determining an ISC according to \eqref{eq:defGama}-\eqref{eq:Phi2def}, the proposed learning-based predictive controller is used to stabilize the system with parameters  $\epsilon_l=0.9$, $\delta=0.1$, $\delta_1=0.1$, and 
\begin{align}\label{eq:paraPC}
	N=5,~Q={\rm diag}\{2,2,2\},~R={\rm diag}\{1,1\}.
\end{align}
 To ensure a fair performance comparison, the objective function of DeePC and RRL is set to be the same as that of the proposed approach.
controller with ITL according to \eqref{eq:costJ}, with matrices $Q$ and $R$ together with the control horizon $N$ selected according to \eqref{eq:paraPC}, and the input and state constraints are set to the same for all three controllers. The regularization weight parameters for DeePC are set to $\lambda_g=30, \lambda_y=10^5$ according to \cite{coulson2019data}, and the length of epochs is set to be the same as the control horizon $N=5$ for RRL according to \cite{umenberger2019robust}. To evaluate the control performance, we define the following weighted square error $J_W$ as
\begin{align}
	J_W=\sum\limits_{k=8}^{200} {\bm x}^{\rm T}(k)Q{\bm x}(k)+{\bm u}^{\rm T}(k)R{\bm u}(k).
\end{align}
The cost is calculated by summing stage costs at time instants $\{8,\ldots,200\}$ since data samples at times instants $\{1,\ldots,7\}$ are used to initialize the ISC.

Other metrics are also used for comparison, including the mean square error (MSE) of the states, the number of open-loop data points required to initiate the predictive controller\footnote{Here $n(k_0)$ is consistent with the notation in Theorem \ref{thm:convlearn}, but it is also abused to represent the number of the data points needed to initialize DeePC and RRL.} ($n(k_0)$), the number of data points used to update the prediction model during closed-loop control ($n(200)$), and the total simulation time (which reflects the computation time used to execute the predictive controllers).

\begin{figure}[htbp]
	\centering
	\includegraphics[width=1\linewidth]{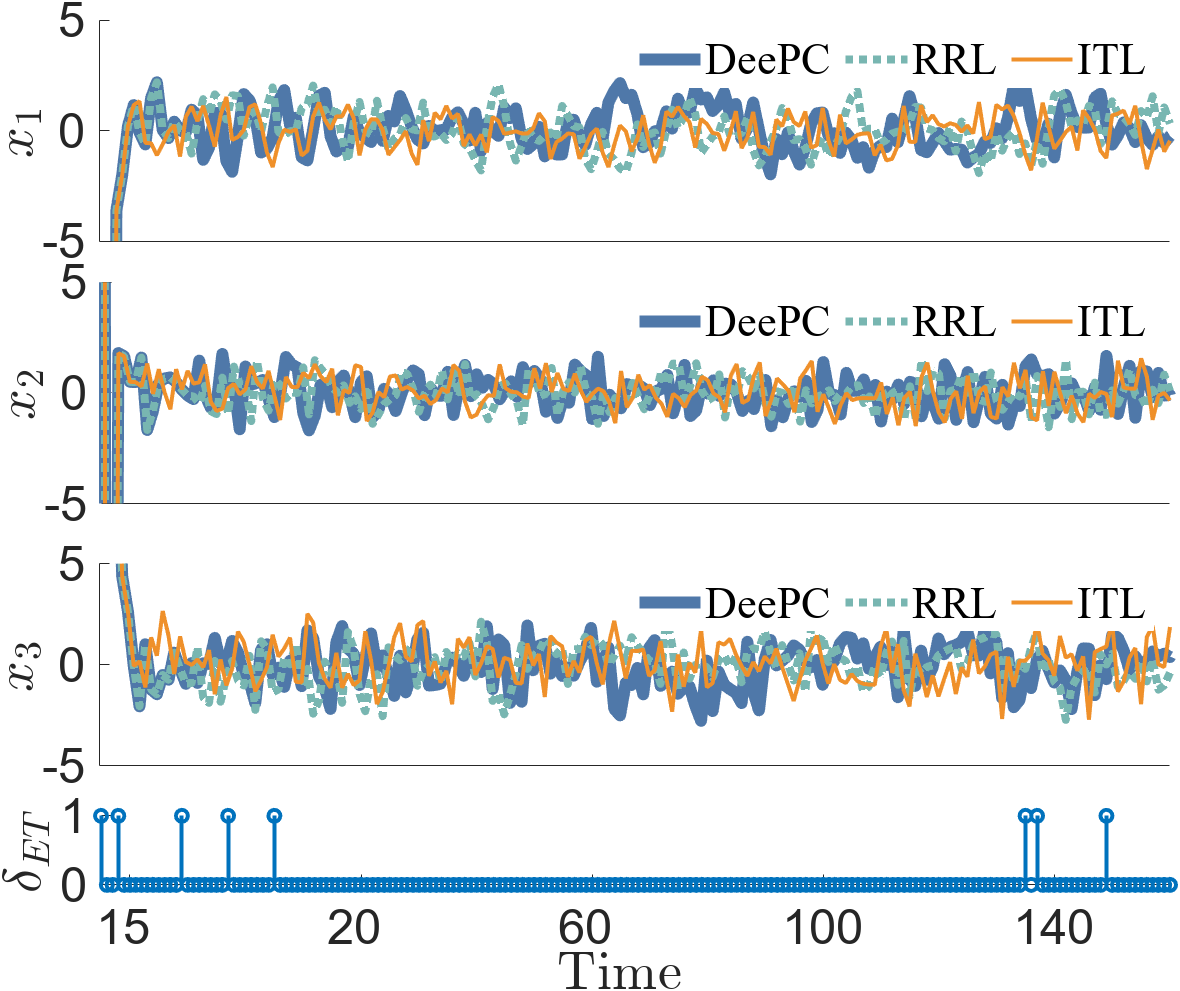}
	\caption{Comparison of control performance between DeePC, RRL, and ITL-based predictive controllers.}
	\label{fig:compare}
\end{figure}

The state trajectories of different controllers are plotted in Fig.~\ref{fig:compare}. The comparison of the performance metrics is provided in Tab.~\ref{tab:1}. The updating instants are marked in the bottom subplot of Fig.~\ref{fig:compare}, where $\delta_{ET}=1$ indicates that a new sample is included in the dataset, and $\delta_{ET}=0$ otherwise. 
In this simulation, to illustrate the effectiveness of the proposed method, the noise is chosen as bounded noise that takes its value from the set $\{-1,0,1\}$.
By observing the comparison results in Fig.~\ref{fig:compare} and Tab.~\ref{tab:1}, the proposed min-max predictive control method with the ITL protocol has a higher data utilization efficiency compared to the DeePC and RRL methods. 
Compared to DeePC, the proposed approach requires less data to enable the initial design of the predictive control (measured by $n(k_0)$). 
This is because the amount of required data is influenced by the prediction length $N$ in DeePC (see Theorem 5.1 in \cite{coulson2019data} for more details).
Compared to RRL, the proposed ITL-based predictive control does not need to update the prediction model frequently (measured by $n(200)$).
As a result, 15 samples are determined to meet the event-triggering condition and thus used for the ITL-based predictive controller. For the ITL-based predictive control, the MSE and cost $J_W$ are smaller than those of other approaches, due to the fact that the proposed method does not rely on assumptions about the noise distribution, thereby exhibiting better robustness against disturbances. Moreover, the average time spent on each data update was 0.04 s, and the average time spent solving the MPC was 0.13 s.

\begin{table}[htbp]
	\centering
	\caption{Comparison of different performance metrics.}
	\label{tab:1}
	\begin{tabular}{cccc}
		\hline
		Method & DeePC [19] & RRL [8] & ITL \\
		\hline
		MSE & 1.34 & 1.27 & \textbf{1.23} \\
		Cost $J_W$ & 1042.32 & 999.30 & \textbf{936.71} \\
		$n(k_0)$ & 23(fixed) & 23 & \textbf{7} \\
		$n(200)$ & 23(fixed) & 198 & \textbf{15}\\ 
		\hline
	\end{tabular}
\end{table}

\vspace{-1em}
\section{Conclusion} \label{sec:conc}
\vspace{-1em}
In this work, we consider a set-membership learning problem for LTI systems subject to disturbances, and introduce an ITL approach to update the learned parametric uncertainty set only when the estimated quantitative incremental information exceeds a threshold. The unknown dynamics of the system are estimated using a high-probability set, and the Lebesgue measure of the set-membership estimate is shown to be exponentially convergent under the proposed ITL mechanism. We also show that the ITL can be integrated with LPC design through utilizing a min-max optimization framework. In our next step, the extension of the proposed results to certain nonlinear dynamic systems will be further explored.

\section*{Acknowledgment}
The authors would like to thank the Associate Editor and the anonymous reviewers for their suggestions which have improved the quality of the work.

\bibliographystyle{plainnat}        
\bibliography{ITL_arxiv}           

\appendix

\section{Details on Solving the  Problem \eqref{eq:MPC1}}\label{App:1-solving}
In summary, the problem \eqref{eq:MPC1} is a min-max optimization problem. The main challenge in solving it lies in ensuring that the constraints hold for all disturbances satisfying $\mathbb{P}({\bm W}(k;N)\in\mathbb{W}_{\delta_1})\geq \delta_1$ and dynamic characteristics \eqref{eq:consAB}, while minimizing the worst-case cost. This appendix will separately discuss the problem reformulation related to disturbances and the methods for handling uncertainties in system parameters.
\subsection{Reformulation w.r.t. Disturbances}
The considered disturbances can be parameterized as 
\begin{align}\label{eq:What}
	\mathbb{W}_{\delta_1}:=\left\{{\bm W}\left|{\rm Tr}\left[\Psi_w\right]\!-\!{\rm Vec}({\bm W})^{\rm T}{\rm Vec}({\bm W})\right.\right\}.
\end{align}
By writing $\mathcal{W}(k)={\rm Vec}({\bm W})$, an equivalent set to $\mathbb{W}_{\delta_1}$ can be defined as
\begin{align}
	\mathbb{W}_{\delta_1}=\left\{\mathcal{W}(k)\left|{\rm Tr}\left[\Psi_w\right]-\mathcal{W}^{\rm T}(k)\mathcal{W}(k)\geq 0\right.\right\},\label{eq:consW2}
\end{align}
with a slight abuse of notation.
Before presenting the reformulation details, a useful lemma is provided below, which can be used to discuss the relationships between different quadratic inequalities.
\begin{lemma}(S-Procedure)\label{lemma:SP}
	Let $F_0, F_1, \ldots, F_k\in\mathbb{R}^{n\times n}$ be symmetric matrices, and $\tau_1, \ldots, \tau_k\geq 0$ be scalars. Then if $F_0\succeq \sum\limits_{i=1}^k\tau_iF_i$, we have
\begin{align}
		\forall \xi\in\mathbb{R}^{n\times 1}:~ \xi^{\rm T}F_1\xi\geq 0,\ldots, \xi^{\rm T}F_k\xi\geq 0 \Rightarrow \xi^{\rm T}F_0\xi\geq 0.
	\end{align} 
\end{lemma}

For simplicity, we write $\mathcal{ W}(k):={\rm Vec}(\hat{\bm W}(k;N))$, $\mathcal{ V}(k):={\rm Vec}({\bm V}(k))$ and $\hat{A}:=A+BK(k)$. Then the states predicted in \eqref{eq:Xpred} can be equivalently rewritten in a matrix form as \eqref{eq:Xeq}.
\begin{figure*}
\begin{align}\label{eq:Xeq}
		\overbrace{\left[\!\!\begin{array}{c}
			\bar{\bm x}(k)\\
			\bar{\bm x}(k+1)\\
			\bar{\bm x}(k+2)\\
			\vdots\\
			\bar{\bm x}(k+N)
		  \end{array}\!\!\right]}^{\bar{X}(k)}
		=&
		\overbrace{\left[\begin{array}{cccc}
			0&\cdots&0&0\\
			B&\cdots&0&0\\
			\hat{A}B&\cdots&0&0\\
			\vdots&\ddots &\vdots&\vdots\\
			\hat{A}^{N-1}B&\cdots& B&0
		  \end{array}\right]}^{\Phi_{AB}}
\overbrace{\left[\begin{array}{c}
{\bm v}(k)\\
\vdots\\
{\bm v}(k\!+\!N\!-\!1)\\
{\bm 0}
	 \end{array}
\right]}^{\mathcal{V}(k)}\notag\\
		   &+
		   \underbrace{\left[\begin{array}{cccc}
			I\\
			\hat{A}\\
			\hat{A}^2\\
			\vdots\\
			\hat{A}^N
		  \end{array}\right]}_{\Phi_{A}}
\bar{\bm x}(k)
+
		   \underbrace{\left[
			\begin{array}{cccc}
				0&0&\cdots&0\\
				I&0&\cdots&0\\
				\hat{A}&I&\cdots&0\\
				\vdots&\vdots&\ddots &\vdots\\
				\hat{A}^{N-1}&\hat{A}^{N-2}&\cdots& I
			\end{array}
		   \right]}_{\Phi_w}
		   \underbrace{\left[
			\begin{array}{c}
				\hat{\bm w}(k)\\
				\hat{\bm w}(k+1)\\
				\vdots\\
				\hat{\bm w}(k\!+\!N\!-\!1)
			\end{array}
		   \right]}_{\mathcal{W}(k)}.
\end{align}
\end{figure*}
Then the cost function \eqref{eq:costJ} can be written in a matrix form as
\begin{align}
	&J_N({\bm x}(k),\mathcal{V}(k),\mathcal{W}(k);A,B)\notag\\
	=&\![\bar{K}(k)\bar{\bm X}(k)\!+\!\bar{\mathcal{V}}(k)]^{\rm T}\bar{R}[\bar{K}(k)\bar{\bm X}(k)\!+\!\bar{\mathcal{V}}(k)]\notag\\
	&\bar{\bm X}(k)^{\rm T}\bar{Q}\bar{\bm X}(k)
	+\notag,
\end{align}
where $\bar{Q}$ and $\bar{R}$ are block diagonal matrices defined as
\begin{align}
	\bar{Q}&:={\rm blkdiag}\{Q,\ldots,Q,P_f\},\\
	\bar{R}&:={\rm blkdiag}\{R,\ldots,R,0\},\\
	\bar{K}(k)&={\rm blkdiag}\{K(k),\ldots,K(k)\}.
\end{align}
Moreover, by defining ${\bm z}(k):=[1~{\bm x}^{\rm T}(k)~\mathcal{W}^{\rm T}(k)]^{\rm T}$, the cost function $J_N({\bm x}(k),\mathcal{V}(k),\mathcal{W}(k);A,B)$ can be written as ${\bm z}^{\rm T}(k)H(A,B,\mathcal{V}(k)){\bm z}(k)$ with $H(A,B,\mathcal{V}(k))$ being a symmetric matrix defined as
\begin{align}
	H(A,B,\mathcal{V}(k)):=
	\left[\begin{array}{ccc}
		H_{11}&H_{12}&H_{13}\\
		H_{12}^{\rm T}&H_{22}&H_{23}\\
		H_{13}^{\rm T}&H_{23}^{\rm T}&H_{33}.
	\end{array}\right]\label{eq:HABV}
	\end{align}
The matrices in \eqref{eq:HABV} are defined as 
\begin{align}
	H_{11}&:=\mathcal{V}^{\rm T}(k)\Phi_{AB}^{\rm T}\hat{Q}(k)\Phi_{AB}\mathcal{V}(k)\notag\\
	&~~~~+\mathcal{V}^{\rm T}(k)\Phi_{AB}^{\rm T}K^{\rm T}(k)R\mathcal{V}(k)\notag\\
	&~~~~+\mathcal{V}^{\rm T}(k)RK(k)\Phi_{AB}\mathcal{V}(k)+\mathcal{V}^{\rm T}(k)R\mathcal{V}(k),\notag\\
	H_{12}&:=\mathcal{V}^{\rm T}(k)\Phi_{AB}^{\rm T}\hat{Q}(k)\Phi_A+\mathcal{V}^{\rm T}(k)RK(k)\Phi_A,\notag\\
	H_{13}&:=\mathcal{V}^{\rm T}(k)\Phi_{AB}^{\rm T}\hat{Q}(k)\Phi_w+\mathcal{V}^{\rm T}(k)RK(k)\Phi_w\notag\\
	H_{22}&:=\Phi_A^{\rm T}\hat{Q}(k)\Phi_A,~H_{23}:=\Phi_{A}^{\rm T}\hat{Q}(k)\Phi_{w},\notag\\
	H_{33}&:=\Phi_{w}^{\rm T}\hat{Q}(k)\Phi_{w}\notag
\end{align}
where $\hat{Q}(k)$ is defined as $\hat{Q}(k):=Q+K^{\rm T}(k)RK(k)$.

Using the notations aforementioned, the following relationship can be obtained:
\begin{align}
	&\forall A,B,\mathcal{V}(k):J_N({\bm x}(k),\mathcal{V}(k),\mathcal{W}(k);A,B)\leq {\bar{V}}\notag\\
	\Longleftrightarrow& {\bm z}^{\rm T}(k)H(A,B,\mathcal{V}(k)){\bm z}(k)\leq {\bm z}^{\rm T}(k)H_{\bar{V}}{\bm z}(k)\notag\\
	\Longleftrightarrow& {\bm z}^{\rm T}(k)(H_{\bar{V}}-H(A,B,\mathcal{V}(k))){\bm z}(k)\geq 0,\label{eq:ineqH}
\end{align}
where $H_{\bar{V}}$ is a symmetric matrix defined as
\begin{align}
	H_{\bar{V}}:=\left[\begin{array}{ccc}
		{\bar{V}}&0&0\\
		0&0&0\\
		0&0&0
	\end{array}\right].
\end{align}

Furthermore, we note that ${\bar{V}}$ is the upper bound of the cost function, which holds for a given state ${\bm x}(k)$ and all disturbances ${\bm W}(k;N)$ satisfying \eqref{eq:consW2}, namely,
\begin{align}
	&\forall {\bm z}(k):~{\bm z}^{\rm T}(k)H_x(k){\bm z}(k)\geq0,~{\bm z}^{\rm T}(k)H_w{\bm z}(k)\geq 0\notag\\
	\Rightarrow &{\bm z}^{\rm T}(k)(H_{\bar{V}}-H(A,B,\mathcal{V}(k))){\bm z}(k)\geq 0,\label{eq:ineqH2}
\end{align}
with $H_w$ and $H_x(k)$ being
\begin{align}
	H_w:=\left[\begin{array}{ccc}
		{\rm Tr}(\Psi_w)&0&0\\
		0&0&0\\
		0&0&-I
	\end{array}\right],~
H_x(k):=\left[\begin{array}{ccc}
		{\bm x}^{\rm T}(k){\bm x}(k)&0&0\\
		0&-I&0\\
		0&0&0
	\end{array}\right].\notag
\end{align}
According to Lemma \ref{lemma:SP}, the relationship in \eqref{eq:ineqH2} holds if 
\begin{align}
\exists \tau_x,\tau_w\geq 0:~H_{\bar{V}}-H(A,B,\mathcal{V}(k))\succeq \tau_xH_x(k)+\tau_wH_w.\label{eq:SP}
\end{align}
For fixed parameters $A,B,\mathcal{V}(k)$, a maximization problem can be reformulated as a minimization problem using inequality \eqref{eq:SP}, namely,
\begin{align}
	&\max J_N({\bm x}(k),\mathcal{V}(k),\mathcal{W}(k);A,B)~~{\rm s.t.}~\mathcal{W}(k)\in\mathbb{W}_{\delta_1}\notag\\
	\Longleftrightarrow& \min\limits_{{\bar{V}}} {\bar{V}}~~{\rm s.t.}~\eqref{eq:SP}~\text{holds}.
\end{align}
Thus the min-max problem \eqref{eq:MPC1} can be reformulated as a min-min problem.

In summary, for given parameters $A,B$, the min-max problem \eqref{eq:MPC1} can be reformulated as a minimization problem:
\begin{align}
	&\min\limits_{\mathcal{V}(k)}\min\limits_{\bar{V}} ~{\bar{V}}\label{eq:MPC2}\\
	&{\rm {s.t.}}~\eqref{eq:SP},\notag\\
	&~~~~\bar{\bm x}(k\!+\!i)=A\bar{\bm x}(k\!+\!i\!-\!1)+\hat{\bm w}(k\!+\!i\!-\!1)\notag\\
		&~~~~~~~~~~~~~~~~~~~+B\mu_c(\bar{\bm x}(k\!+\!i\!-\!1),{\bm v}(k\!+\!i\!-\!1)),\notag\\
		&~~~~\mu_c(\bar{\bm x}(k\!+\!i),{\bm v}(k\!+\!i))=K(k)\bar{\bm x}(k\!+\!i)+{\bm v}(k\!+\!i),\notag\\
		&~~~~\mu_c(\bar{\bm x}(k+i),{\bm v}(k+i))\in\mathbb{U},~\bar{\bm x}(k+i)\in\mathbb{X}.\notag
\end{align}

Up to now, the maximization problem have been reformulated as a minimization problem with LMIs. The reformulation w.r.t. disturbances can be achieved thanks to the linear relationship between the state and noise, which allows us to use LMIs to describe state constraints, input constraints, and the maximum value of the cost function. However, the cost function and constraints still depend nonlinearly on the dynamic characteristics $(A,B)$, which are more challenging to handle and will be discussed in the next section.

\subsection{Handling Uncertain Parameters}
The system parameters $(A,B)$ are uncertain and can be described by the set $\Gamma(\Psi(n(k),\delta),\mathcal{D}(k))$. The main challenge in solving the reformulated problem \eqref{eq:MPC2} lies in the nonlinearity of the cost function and constraints w.r.t. $(A,B)$. To address this issue, we use the scenario approach \citep{1632303} to handle the uncertainties in system parameters. The scenario approach is a powerful method for dealing with uncertainties in optimization problems \citep{micheli2022scenario}, which allows us to approximate the uncertain set by a finite number of scenarios.

In the scenario approach, we sample a finite number ($N_s$ in this work) of scenarios from the uncertain set $\Gamma(\Psi(n(k),\delta),\mathcal{D}(k))$. For each scenario, we can obtain a specific pair $(A^{(i)},B^{(i)})$, which can be used to evaluate the cost function and constraints. The key idea is to ensure that the constraints hold for all sampled scenarios.
Let $F_P(A,B,\mathcal{V})$ be an indicator function that specifies whether the problem \eqref{eq:MPC2} is feasible for a given pair $(A,B)$ and $\mathcal{V}$, i.e.,
\begin{align}
	&F_P(A,B,\mathcal{V})\notag\\
	=&\left\{\begin{array}{ll}
		1, & \text{if problem \eqref{eq:MPC2} is feasible for}~(A,B)~\text{and}~\mathcal{V},\\
		0, & \text{otherwise.}
	\end{array}\right.\notag
\end{align}

For convenience, some concepts are defined below, which can be referred to in \citet{1632303}.
\begin{define}(Probability of Violation~\citep{1632303}):~For a given parameter $(A,B)$, the probability of violation $P_V(A,B)$ is defined as
	\begin{align}
		&P_V(\mathcal{V})\notag\\
		=&\mathbb{P}\left(F_P(A,B,\mathcal{V})=0|(A,B)\in \Gamma(\Psi(n(k),\delta),\mathcal{D}(k))\right).\notag
	\end{align}
\end{define}
\begin{define}($\epsilon_s$-Level Solution~\citep{1632303}):~Let $\epsilon_s\in(0,1)$. We say that $\mathcal{V}$ is an $\epsilon_s$-level robustly feasible solution (or, more simply, an $\epsilon_s$-solution), if $P_V(\mathcal{V})\leq \epsilon_s$.
	\end{define}

Let $(A^{(i)},B^{(i)}), i\in\{1,\ldots,N_s\}$ be $N_s$ independent identically distributed samples from the uncertain set $\Gamma(\Psi(n(k),\delta),\mathcal{D}(k))$, which can be uniformly sampled from the set $\Gamma(\Psi(n(k),\delta),\mathcal{D}(k))$. The set ${\Gamma}({\Psi}(n(k),\delta),\mathcal{D}(k))$ can be parameterized as:
\begin{align}
	[A~B]=[\hat{A}~\hat{B}]+L_{\tilde{\Phi}_1}C_{\rm AB}R^{\rm T}_{\Phi_2(\mathcal{D}(k))},\notag
\end{align}
where $L_{\tilde{\Phi}_1}$ is the lower triangular matrix obtained from the Cholesky decomposition of $\tilde{\Phi}_1$ (i.e., $\tilde{\Phi}_1=L_{\tilde{\Phi}_1}L_{\tilde{\Phi}_1}^{\rm T}$), $R^{\rm T}_{\Phi_2(\mathcal{D}(k))}$ is the upper triangular matrix from the Cholesky decomposition of $\Phi_2(\mathcal{D}(k))$, and $C_{\rm AB}$ is a parameter constrained by $\|C_{\rm AB}\|_2\leq 1$.

Then the probability of violation can be discussed using the sampled scenarios as follows.

\begin{proposition}
	For $\epsilon_s,\beta_s\in(0,1)$, if $N_s$ problems are feasible for the sampled scenarios $(A^{(i)},B^{(i)})$ and $\mathcal{V}$, and
		\begin{align}
		N_s\geq\left\lceil \frac{2}{\epsilon_s}\ln\frac{1}{\beta_s}+2n_{\mathcal{V}}+\frac{2n_{\mathcal{V}}}{\epsilon_s}\ln\frac{2}{\epsilon_s}\right\rceil, 
		\end{align}
		then the solution $\mathcal{V}$ is an $\epsilon_s$-level robustly feasible solution for the problem \eqref{eq:MPC2} with probability at least $1-\beta_s$, where $n_{\mathcal{V}}=n_u\times N$ and $\lceil\cdot\rceil$ denotes the ceiling function.
\end{proposition}
\begin{pf}
The proof can be found in \citet{1632303} and thus is omitted here.\hfill\QEDopen
\end{pf}

Furthermore, a disturbance invariant set $\mathbb{Z}_{I}(k)$ is calculated according to Mayne et al. [2005], which satisfies
\begin{align}
	&(A+BK(k))\mathbb{Z}_I(k)\oplus\mathbb{W}_{\delta_1}\subseteq\mathbb{Z}_I(k),\notag\\
	&~~~~~~~(A,B)\in\Gamma\left(\Psi(n(k),\delta),\mathcal{D}(k)\right),\notag
\end{align}
where $\oplus$ denotes the Minkowski sum. Using the disturbance invariant set $\mathbb{Z}_I(k)$, the tightened state constraints $\bar{\mathbb{X}}(k)$ and input constraints $\bar{\mathbb{U}}(k)$ can be obtained as 
\begin{align}
	\bar{\mathbb{U}}(k):=\mathbb{U}\ominus K(k)\mathbb{Z}_I(k),~\bar{\mathbb{X}}(k):=\mathbb{X}\ominus \mathbb{Z}_I(k),\notag
\end{align}
where $\ominus$ denotes the Minkowski difference. These tightened constraints ensure satisfaction of the original constraints $\mathbb{X}$ and $\mathbb{U}$ for all disturbances in $\mathbb{W}_{\delta_1}$:
\begin{align}
	&\forall \bar{\bm x}\in\bar{\mathbb{X}}(k), \forall{\bm w}\in\mathbb{W}_{\delta_1}\Rightarrow\bar{\bm x}+{\bm w}\in\mathbb{X};\notag\\
	&\forall \!K(k)\bar{\bm x}\!+\!{\bm v}\!\in\!\bar{\mathbb{U}}(k), \forall{\bm w}\!\in\!\mathbb{W}_{\delta_1}\!\Rightarrow\!K(k)[\bar{\bm x}\!+\!{\bm w}]\!+\!{\bm v}\!\in\!\mathbb{U}.\notag
\end{align}
{Based on these discussions, the problem \eqref{eq:MPC2} has been} reformulated as a minimization problem with LMIs, and the uncertainties in system parameters have been handled using the scenario approach:
\begin{align}
	&\min\limits_{\mathcal{V}(k)}\min\limits_{\bar{V}} ~{\bar{V}}\label{eq:MPC3}\\
		&{\rm {s.t.}}~\eqref{eq:SP},\notag\\
		&~~~~\bar{\bm x}(k\!+\!i)=A^{(j)}\bar{\bm x}(k\!+\!i\!-\!1)+B^{(j)}\mu_c(\bar{\bm x}(k\!+\!i\!-\!1),{\bm v}(k\!+\!i\!-\!1)),~j\in\{1,\ldots,N_s\},\notag\\
			&~~~~\mu_c(\bar{\bm x}(k\!+\!i),{\bm v}(k\!+\!i))=K(k)\bar{\bm x}(k\!+\!i)+{\bm v}(k\!+\!i),\notag\\
			&~~~~\mu_c(\bar{\bm x}(k+i),{\bm v}(k+i))\in\bar{\mathbb{U}}(k),~\bar{\bm x}(k+i)\in\bar{\mathbb{X}}(k).\text{''}\notag
\end{align}

\section{Proof of Proposition \ref{prop:lowbounddelta}}\label{app:prop1}
To facilitate the description of the proof, we rewrite the matrices involved in inequality \eqref{eq:Deltaxi} in the form of block matrices as follows:
	\begin{align}
	\left[\!\!
	\begin{array}{c:c}
	M_{11} & M_{12} \\ \hdashline
	M_{21} & M_{22}
	\end{array}\!\!\right]\!\!&:=\!\!\left[\begin{array}{c:c}
		P_f^{-1}(k+1)&{\bm 0}\\\hdashline
		{\bm 0}&-\left[\begin{array}{c}
			I\\K(k)
		\end{array}\right][I~K^{\rm T}(k)]
	\end{array}\right]\notag
	\end{align}
	\begin{align}
	\left[\!\!
	\begin{array}{c:c}
	N_{11} & N_{12} \\ \hdashline
	N_{21} & N_{22}
	\end{array}\!\!\right]\!\!&:=\!\!\Xi(k)\tilde{\Psi}(n(k))\Xi^{\rm T}(k),
	\end{align}
	where $N_{11}=\Phi_1(n(k),\delta)\!-\!X_+(k)X_+^{\rm }(k)$, and $N_{22}=-\left[X_-(k)^{\rm T}~U_-(k)^{\rm T}\right]^{\rm T}\star^{\rm T}$.
	Then, according to the Schur complement theorem, the following inequalities can be obtained from \eqref{eq:Deltaxi}:
	\begin{align}
		&M_{11}-\xi N_{11}\succeq 0 \label{eq:ieq11},\\
		&\xi N_{22}-\frac{1}{\delta_M} M_{22}\succeq 0 \label{eq:ieq22}.
	\end{align}
	
	For symmetric positive definite matrices $M_{11}, N_{11}$, we have 
	\begin{align}
		\xi\leq\hat{\lambda}_{\rm min}=\lambda_{\rm min}(M_{11},N_{11})\leq \frac{{\bm v}^{\rm T}M_{11}{\bm v}}{\bm {v}^{\rm T}N_{11}{\bm v}},\label{eq:xihatl}
	\end{align}
	where the last inequality is obtained according to the property of Rayleigh quotient. It can be observed from \eqref{eq:xihatl} that the inequality $\xi\leq\hat{\lambda}_{\rm min}$ is the sufficient condition of the inequality \eqref{eq:ieq11}.
	
	Similarly, a sufficient condition for the inequality \eqref{eq:ieq22} can be obtained as 
	\begin{align}
		\frac{1}{\xi\delta_M}\leq\check{\lambda}_{\rm min}=\lambda_{\rm min}(N_{22},M_{22})\label{eq:xicheck}.
	\end{align}
	
	By combining inequalities in \eqref{eq:xihatl} and \eqref{eq:xicheck}, we have
	\begin{align}
		\delta_M\geq\frac{1}{\xi\check{\lambda}_{\rm min}}\geq \frac{1}{{\hat{\lambda}_{\rm min} \check{\lambda}_{\rm min}}},\notag
	\end{align}
	which completes the proof. 

\section{Proof of Proposition \ref{prop:feas}}\label{app:feas}
To prove the recursive feasibility of the MPC problem \eqref{eq:MPC1}, we assume that the considered problem is feasible at time instant $k$ and aim to prove its feasibility at time instant $k+1$.

		Let $V^*(k)$ be the optimal solution of the problem \eqref{eq:MPC1}. For parameters $(A,B)\in\bar{\Gamma}(\Psi(\bar{n}(k),\delta),\bar{\mathcal{D}}(k))$ and distrubances $\hat{W}(k;N)=[\hat{\bm w}(k),\ldots,\hat{\bm w}(k+N-1)]\in\mathbb{W}_{\delta_1}^N$, the possible states are denoted for $i\in\{1,\ldots,N\}$
		\begin{align}
		\hat{\bm x}(k\!+\!i|k)&\!=\!A\hat{\bm x}(k\!+\!i\!-\!1|k)\!+\!B{\bm u}^*(k\!+\!i\!-\!1)\!+\!\hat{\bm w}(k\!+\!i\!-\!1),\\
		\hat{\bm x}(k|k)&={\bm x}(k|k),
		\end{align}
		where ${\bm u}^*(k+i)=K\hat{\bm x}(k+i)+{\bm v}^*(k+i), i\in\{0,\ldots,N-1\}$ with ${V}^*(k)=\{{\bm v}^*(k),\ldots,{\bm v}^*(k+N-1)\}$ being the optimal solution of the problem \eqref{eq:MPC1}.
		
		The possible state sets are written as 
		\begin{align}
		&\hat{\mathbb{X}}(k+i|k)\notag\\
		:=&\{\hat{\bm x}(k+i|k)\big|(A,B)\in\Gamma(\Psi(n(k),\delta),\mathcal{D}(k)),\notag\\
		&~~~~~~~~~~~~~~~~~~\hat{W}(k;N)\in\mathbb{W}_{\delta_1}^N\},i\in\{1,\ldots,N\}.\notag
		\end{align}
		The feasibility of the constrained optimization problem \eqref{eq:MPC1} leads to 
		\begin{align}
		\hat{\bm x}(k+i|k)&\in \hat{\mathbb{X}}(k+i|k)\subseteq\mathbb{X},~i\in\{1,\ldots,N-1\},\notag\\
		\hat{\bm x}(k+N|k)&\in\hat{\mathbb{X}}(k+N|k)\subseteq\mathbb{X}_f(k).\notag
		\end{align}
		
		Then the feasibility of the problem \eqref{eq:MPC1} at time instant $k+1$ can be discussed. Consider a sequence $V(k+1)=\{{\bm v}^*(k+1),\ldots,{\bm v}^*(k+N-1),{\bm 0}\}$ and the corresponding control sequence $\hat{U}(k+1)=\{\hat{\bm u}(k+1),\ldots, \hat{\bm u}(k+N)\}$. By recalling that $\bar{\Gamma}(\Psi(\bar{n}(k+1),\delta),\bar{\mathcal{D}}(k+1))\subset\bar{\Gamma}(\Psi(\bar{n}(k),\delta),\bar{\mathcal{D}}(k))$, we have
		\begin{align}
			{\bm x}(k+1)&\in \hat{\mathbb{X}}(k+1|k),\notag\\
			\hat{\mathbb{X}}(k+i|k+1)&\subset\hat{\mathbb{X}}(k+i-1|k)\subseteq\mathbb{X},~i\in\{1,\ldots,N\}.\notag
		\end{align}
		
		Moreover, the terminal constraint $\mathbb{X}_f(k)$ is a control invariant set according to inequality \eqref{eq:cond1}, and thus the terminal state satisfies 
		\begin{align}
			\hat{\mathbb{X}}(k+N+1|k+1)\subset \mathbb{X}_f.\notag
		\end{align}
		which ensures the recursive feasibility of the proposed ITL-based MPC.



\end{document}